\documentclass[12pt,a4paper]{article}
\usepackage[english]{babel}
\usepackage{amsmath,amssymb}
\usepackage[dvips]{graphicx}
\usepackage{amsfonts}
\usepackage{bbm}
\usepackage{cite}

\def\baselinestretch{1.27}%

\newcommand{\be}{\begin{equation}}
\newcommand{\ee}{\end{equation}}
\newcommand{\bea}{\begin{eqnarray}}
\newcommand{\eea}{\end{eqnarray}}

\newcommand {\ignore}[1]{}
\newcommand{\lsim}{
\mathrel{\hbox{\rlap{\hbox{\lower4pt\hbox{$\sim$}}}\hbox{$<$}}}}
\newcommand{\gsim}{
\mathrel{\hbox{\rlap{\hbox{\lower4pt\hbox{$\sim$}}}\hbox{$>$}}}}

\let\vev\VEV
\def\ZP{$Z^\prime$ }
\def\e6{$E(6)$}
\def\10{$SO(10)$}
\def\21{$SU(2) \otimes U(1) $}
\def\lr{$SU(3) \otimes SU(2)_L \otimes SU(2)_R \otimes U(1)_{B-L}$ }
\def\422{$SU(4) \otimes SU(2) \otimes SU(2)$ }
\def\321{$SU(3) \otimes SU(2) \otimes U(1)$ }
\def\O{\hbox{$\cal O$ }}
\def\meff{\langle m_{\nu} \rangle}
\def\lfv{lepton flavour violation }
\def\lnv{lepton number violation }
\newcommand{\ed}{\end{document}}
\DeclareMathAlphabet{\mathsc}{OT1}{cmr}{m}{sc}

\newcommand{\CL}   {C.L.}
\newcommand{\dof}  {d.o.f.}

\newcommand{\eVq}  {\rm{eV}^2}
\newcommand{\Sol}  {\mathsc{sol}}
\newcommand{\Atm}  {\mathsc{atm}}

\newcommand{\Dms}  {\Delta m^2_\Sol}
\newcommand{\Dma}  {\Delta m^2_\Atm}

\def \znbb {$0\nu\beta\beta$ }
\def \nbb {$\beta\beta_{0\nu}$ }

\def\meff{\langle m_{\nu} \rangle}

\newcommand{\AddrAHEP}{%
 AHEP Group, Instituto de F\'{\i}sica Corpuscular,
  C.S.I.C. -- Universitat de Val{\`e}ncia \\
  Edificio de Institutos de Paterna, Apartado 22085,
  E--46071 Val{\`e}ncia, Spain\\}

\begin{document}
\begin{center}
\begin{Large}
{\bf  Neutrino physics overview}

\end{Large}
\vspace{0.5cm}
{J. W. F. Valle \\
 \AddrAHEP}
\end{center}
\begin{abstract} 
  Seesaw-type and low-scale models of neutrino masses are reviewed,
  along with the corresponding structure of the lepton mixing matrix.
  The status of neutrino oscillation parameters as of June 2006 is
  given, including recent fluxes, as well as latest SNO, K2K and MINOS
  results.
  Some prospects for the next generation of experiments are given.
  This writeup updates the material presented in my lectures at the
  Corfu Summer Institute on Elementary Particle Physics in September
  2005.
\end{abstract} 

\tableofcontents

\section{Introduction}
\label{sec:introduction}

The historic discovery of neutrino
oscillations~\cite{fukuda:2002pe,ahmad:2002jz,araki:2004mb,Kajita:2004ga,ahn:2002up}
marks a turning point in particle and nuclear physics and implies that
neutrinos have mass. This possibility has been first suggested by
theory since the early eighties, both on general grounds and on the
basis of different versions of the seesaw
mechanism~\cite{Minkowski:1977sc,Orloff:2005nu,Weinberg:1980bf,schechter:1980gr,schechter:1982cv,Lazarides:1980nt}.

The general characterization of neutrino mass theories in \321 terms
provided a model-independent basis to analyse the
seesaw~\cite{schechter:1980gr,schechter:1982cv}. It also indicated a
fundamental difference between the lepton and the quark mixing
matrices, namely, the appearance of new phases associated to the
Majorana nature of neutrinos~\cite{schechter:1980gr}.

Irrespective of what the ultimate origin of neutrino mass may turn out
to be, the basic gauge theoretic mechanism to account for the
smallness of neutrino mass is in terms of the feebleness of B-L
violation. The seesaw is one realization of the idea, by far not
unique. There are two classes of theories of neutrino mass, that
differ by the scale at which L-symmetry is broken. They are summarized
in Sec.~\ref{sec:origin-neutrino-mass}.
The corresponding structure of the lepton mixing matrix that follows
from theory is described Sec.~\ref{sec:lepton-mixing-matrix}. This
forms the basis for the analysis of the data from current neutrino
oscillation
experiments~\cite{fukuda:2002pe,ahmad:2002jz,araki:2004mb,Kajita:2004ga,ahn:2002up}~\cite{apollonio:1999ae,boehm:2001ik}.
The status of neutrino mass and mixing parameters as determined from
the world's neutrino oscillation data within the simplest
CP-conserving three-neutrino mixing scheme is summarized in
Sec.~\ref{sec:stat-neutr-oscill}~\cite{Maltoni:2004ei}.
In addition a determination of the solar angle $\theta_{12}$, the
atmospheric angle $\theta_{23}$ and the corresponding mass squared
splittings $\Dms$ and $\Dma$, one gets a constraint on the last angle
in the three--neutrino leptonic mixing matrix, $\theta_{13}$.
Together with the small ratio $\Dms/\Dma$ the angle $\theta_{13}$
holds the key for further progress in neutrino oscillation searches.

Some attempts at predicting neutrino masses and mixing are given in
Sec.~\ref{sec:pred-neutr-mixing}.
Lepton number violating processes such as neutrinoless double beta
decay~\cite{elliott:2002xe,doi:1985dx} are briefly discussed in
Sec.~\ref{sec:neutr-double-beta}.
Searching for \znbb constitutes a very important goal for the future,
as this will probe the fundamental nature of neutrinos, irrespective
of the process that induces it, a statement known as the ``black-box''
theorem~\cite{Schechter:1982bd}.  In addition, \znbb will be sensitive
to the absolute scale of neutrino mass and to CP violation induced by
the so-called Majorana phases~\cite{schechter:1980gr}, inaccessible in
conventional
oscillations~\cite{bilenky:1980cx,Schechter:1981gk,doi:1981yb}.
Finally in Sec.~\ref{sec:non-stand-neutr} the robustness of the
oscillation interpretation and the role of non-standard neutrino
interactions in future precision oscillation studies is briefly
mentioned.

\section{Dirac and Majorana masses}
\label{sec:dirac-major-mass}

Electrically charged fermions must be Dirac type. In contrast,
electrically neutral fermions, like neutrinos (or supersymmetric ``
inos''), are expected to be Majorana-type on general grounds,
irrespective of how they acquire their mass.  Phenomenological
differences between Dirac and Majorana neutrinos are tiny for most
processes, such as neutrino oscillations: first because neutrinos are
known to be light and, second, because the weak interaction is chiral,
well described by the V-A form.  

The most basic spin $1/2$ fermion corresponding to the lowest
representation of the Lorentz group is given in terms of a 2-component
spinor $\rho$, with the following free
Lagrangean~\cite{schechter:1980gr}
\begin{equation}
{\cal L}_M=-i\rho^{\dagger} \sigma_{\mu} \partial_{\mu} \rho
-\frac{m}{2}\rho^T \sigma_2 \rho + H.C.
\label{eq:LM}
\end{equation}
where $\sigma_i$ are the usual Pauli matrices and $\sigma_4=-i \: I$,
$I$ being the $2 \times 2$ identity matrix. I use Pauli's metric
conventions, where $a.b \equiv a_{\mu} b_{\mu} \equiv \vec{a} \cdot
\vec{b} + a_4 b_4$, $a_4=ia_0$.  Under a Lorentz transformation, $x
\to \Lambda x$, the spinor $\rho$ transforms as $\rho \to
S(\Lambda)\rho(\Lambda^{-1}x)$ where $S$ obeys
\begin{equation}
S^{\dagger} \sigma_{\mu} S = \Lambda_{\mu \nu}\sigma_{\nu}
\label{eq:HOMO}
\end{equation}
The kinetic term in Eq.~(\ref{eq:LM}) is clearly invariant, and so is
the mass term, as a result of unimodular property $det\:S=1$.
However, the mass term is not invariant under a phase transformation
\begin{equation}
\rho \to  e^{i \alpha} \rho
\label{eq:repha0}
\end{equation}
The equation of motion following from Eq.~(\ref{eq:LM}) is
\begin{equation}
-i \sigma_{\mu} \partial_{\mu} \rho = m \sigma_2 \rho^*
\label{eq:EQM}
\end{equation}
As a result of the conjugation and Clifford properties of the
$\sigma$-matrices, one can verify that each component of the spinor
$\rho$ obeys the Klein-Gordon wave-equation.
 
Start from the usual Lagrangean describing of a massive spin $1/2$
Dirac fermion, given as
\begin{equation}
{\cal L}_D= - \bar{\Psi} \gamma_{\mu} \partial_{\mu} \Psi -
m\:\bar{\Psi} \Psi,
\label{eq:LD}
\end{equation}
where by convenience we use the chiral representation of the Dirac
algebra $\gamma_{\mu} \gamma_{\nu} + \gamma_{\nu} \gamma_{\mu} =
2\:\delta_{\mu \nu}$ in which $\gamma_5$ is diagonal, \bea \gamma_i =
\left(\begin{array}{ccccc}
    0 & -i\sigma_i\\
    i\sigma_i & 0\\
\end{array}\right)\: &
\gamma_4 = \left(\begin{array}{ccccc}
                        0 & I\\
                        I & 0\\
\end{array}\right)&
\gamma_5 = \left(\begin{array}{ccccc}
                        I & 0\\
                        0 & -I\\
\end{array}\right)\,.
\label{eq:GAMMAS}
\eea
In this representation the charge conjugation matrix $C$ obeying
\bea
C^T = - C\\
C^\dagger = C^{-1}\\
C^{-1}\:\gamma_{\mu}\:C = -\:\gamma_{\mu}^T
\label{eq:conj}
\eea is simply given in terms of the basic conjugation matrix
$\sigma_2$ as
\begin{equation}
C = \left(\begin{array}{ccccc}
                        -\sigma_2 & 0\\
                        0 & \sigma_2 \\
\end{array}\right)\,.
\label{eq:CHCONJ}
\end{equation}
In order to display clearly the relationship between the Majorana
theory in Eq.~(\ref{eq:LM}) and the familiar Dirac Lagrangean in
Eq.~(\ref{eq:LD}), one splits a Dirac spinor as
\begin{equation}
\Psi_D = \left(\begin{array}{ccccc}
                \chi\\
                \sigma_2\:\phi^*\\
                \end{array}\right)\,,
\label{eq:PSID}
\end{equation}
so that the corresponding charge-conjugate spinor
$\Psi_D^c = C\:\bar{\Psi}_D^T$ is the same as $\Psi_D$ but
exchanging $\phi$ and $\chi$, i.~e.
\begin{equation}
\Psi_D^c = \left(\begin{array}{ccccc}
                \phi\\
                \sigma_2\:\chi^*\\
                \end{array}\right)\,.
\label{eq:PSIDCONJ}
\end{equation}
A 4-component spinor is said to be Majorana or self-conjugate if $\Psi
= C \bar{\Psi}^T$ which amounts to setting $\chi = \phi$.  Using
Eq.~(\ref{eq:PSID}) we can rewrite Eq.~(\ref{eq:LD}) as
\begin{equation}
{\cal L}_D=-i\sum_{\alpha=1}^2 \rho_{\alpha}^{\dagger} \sigma_{\mu}
\partial_{\mu} \rho_{\alpha}
-\frac{m}{2} \sum_{\alpha=1}^2 \rho_{\alpha}^T \sigma_2 \rho_{\alpha} + H.C.
\label{eq:LD2}
\end{equation}
where
\bea
\chi = \frac{1}{\sqrt2} (\rho_2 + i\rho_1) \nonumber \\
\phi = \frac{1}{\sqrt2} (\rho_2 - i\rho_1)
\label{eq:DECOMP}
\eea are the left handed components of $\Psi_D$ and of the
charge-conjugate field $\Psi_D^c$, respectively. This way the Dirac
fermion is shown to be equivalent to two Majorana fermions of equal
mass. The $U(1)$ symmetry of the theory described by Eq.~(\ref{eq:LD})
under $\Psi_D \to e^{i \alpha} \Psi_D$ corresponds to continuous
rotation symmetry between $\rho_1$ and $\rho_2$
\begin{eqnarray*}
\rho_1 \to \cos \theta \rho_1 + \sin \theta \rho_2 \\
\rho_2 \to -\sin \theta \rho_1 + \cos \theta \rho_2
\label{eq:repha1}
\end{eqnarray*}
which result from the mass degeneracy between the $\rho$'s, showing
that, indeed, the concept of fermion number is not basic.

The mass term in Eq.~(\ref{eq:LM}) vanishes unless $\rho$ and $\rho^*$
are anti-commuting, so the Majorana fermion is, right from the start,
a quantized field. The solutions to Eq.~(\ref{eq:LM}) are easily
obtained in terms of those of Eq.~(\ref{eq:LD}), which are well known:
\bea
\Psi_M = (2\pi)^{-3/2} \int d^3k \sum_{r=1}^2 (\frac{m}{E})^{1/2} [
e^{i k.x} A_r(k) u_{Lr}(k) + e^{-i k.x} A_r^{\dagger}(k) v_{Lr}(k)] ,
\label{eq:PSIM}
\eea 
where $u=C\:\bar{v}^T$ and $E(k) = (\vec{k}^2 + m^2)^{1/2}$ is the
mass-shell condition.  The creation and annihilation operators obey
canonical anti-commutation rules and, like the $u$'s and $v$'s, depend
on the momentum $k$ and helicity label $r$. 
The expression in Eq.~(\ref{eq:PSIM}) describes the basic Fourier
expansion of a massive Majorana fermion. It differs from the usual
Fourier expansion for the Dirac spinor in Eq.~(\ref{eq:PSID2}) in two
ways:
\begin{itemize}
\item spinors are two-component, as there is a chiral projection on
  the $u$'s and $v$'s
\item there is only one Fock space, particle and anti-particle
  coincide, showing that a massive Majorana fermion corresponds to one
  half of a conventional massive Dirac fermion.
\end{itemize}
The $u$'s and $v$'s are the same wave functions that appear in the
Fourier decomposition the Dirac field
\bea 
\Psi_D = (2\pi)^{-3/2} \int d^3k
\sum_{r=1}^2 (\frac{m}{E})^{1/2} [ e^{i k.x} a_r(k) u_r(k) + e^{-i
  k.x} b_r^{\dagger}(k) v_r(k)]\;.
\label{eq:PSID2}
\eea

Using the helicity eigenstate wave-functions,
\bea \vec{\sigma} \cdot \vec{k} \: u_L^{\pm}(k) = \pm \mid \vec{k}
\mid u_{L}^{\pm}(k)\\
\vec{\sigma} \cdot \vec{k} \: v_L^{\pm}(k) = \mp \mid \vec{k} \mid
v_{L}^{\pm}(k)
\label{eq:limit}
\eea 
one can show that, out of the $4$ linearly independent wave functions
$u_{L}^{\pm}(k)$ and $v_{L}^{\pm}(k)$, only two survive as the mass
approaches zero, namely, $u_{L}^{-}(k)$ and $v_{L}^{+}(k)$
~\cite{schechter:1981hw}. This way the Lee-Yang two-component massless
neutrino theory is recovered as the massless limit of the Majorana
theory.

Two independent propagators follow from Eq.~(\ref{eq:LM}),
\bea
<0 \mid \rho(x) \: \rho^*(y) \mid 0>
=i \sigma_{\mu} \partial_{\mu} \Delta_F(x-y;m)\\
\label{eq:NORMAL}
<0 \mid \rho(x) \: \rho (y) \mid 0>
= m \: \sigma_2 \: \Delta_F(x-y;m)
\label{eq:LNV}
\eea 
where $\Delta_F(x-y;m)$ is the usual Feynman function.  The first one
is the ``normal'' propagator that intervenes in total lepton number
conserving ($\Delta L = 0 $) processes, while the one in
Eq.~(\ref{eq:LNV}) describes the virtual propagation of Majorana
neutrinos in $\Delta L = 2 $ processes such as neutrinoless
double-beta decay.
 
The Lagrangean in Eq.~(\ref{eq:LM}) can easily be generalized to an
arbitrary number of Majorana neutrinos, giving
\begin{equation}
L_M=-i \sum_{\alpha=1}^{n} \rho_{\alpha}^{\dagger} \sigma_{\mu} \partial_{\mu} \rho_{\alpha}
-\frac{1}{2} \sum_{\alpha,\beta=1}^{n} M_{\alpha\beta}\rho_{\alpha}^T \sigma_2 \rho_{\beta} + H.C.
\label{eq:LM2}
\end{equation}
where the sum runs over the ``neutrino-type'' indices $\alpha$ and
$\beta$. By Fermi statistics the mass coefficients $M_{\alpha \beta}$
must form a symmetric matrix, in general complex. This matrix can
always be diagonalized by a complex $n \times n$ unitary matrix
$U$~(See~\cite{schechter:1980gr} for the proof)
\begin{equation}
M_{diag} = U^T M U \:.
\end{equation}
When $M$ is real its diagonalizing matrix $U$ may be chosen to be
orthogonal and, in general, the mass eigenvalues can have different
signs. These may be assembled as a signature matrix
\begin{equation}
\eta = diag(+,+,...,-,-,..)
\end{equation}
For two neutrino types there are two classes of models, one with $\eta
= diag(+,-)$ and another characterized by $\eta = diag(+,+)$.  The
class with $\eta = diag(+,-)$ contains as a limit the case where the
two fermions make up a Dirac neutrino.
Note that one can always make all masses positive by introducing
appropriate phase factors in the wave functions, such as the factors
of $i$ in Eq.~(\ref{eq:DECOMP}).
However, when interactions are added these signs become physical. As
emphasized by Wolfenstein, they play an important role in the
discussion of \znbb (neutrinoless double beta
decay)~\cite{Wolfenstein:1981rk}.
  
\section{The origin of neutrino mass}
\label{sec:origin-neutrino-mass}

\begin{table}
  \centering
\begin{math}
\begin{array}{|c|c|} \hline
& \ \ \ {\mbox SU(3)\otimes SU(2)\otimes U(1)} \\
\hline
L_a = (\nu_a, l_a)^T & (1,2,-1)\\
e_a^c   & (1,1,2)\\
\hline
Q_a = (u_a, d_a)^T    & (3,2,1/3)\\
u_a^c   & (\bar{3},1,-4/3)\\
d_a^c   & (\bar{3},1,2/3)\\
\hline
\Phi  & (1,2,1)\\
\hline
\end{array}
\end{math}
\caption{Matter and scalar multiplets of the Standard Model (SM)}
\label{tab:sm}
\end{table}

Table \ref{tab:sm} gives the fifteen basic building blocks of matter.
They are all 2-component sequential ``left-handed'' chiral fermions,
one set for each generation.  Parity violation in the weak interaction
is incorporated ``effectively'' by having ``left'' and ``right''
fermions transform differently with respect to the \321 gauge group.
In contrast to charged fermions, neutrinos come only in one chiral
species.

It has been long noted by Weinberg~\cite{Weinberg:1980bf} that one can
add to the Standard \321 Model (SM) an effective dimension-five
operator $\O = \lambda L \Phi L \Phi$ where $L$ denotes a
lepton doublet for each generation and $\Phi$ is the SM scalar doublet.
\begin{figure}[h] \centering
    \includegraphics[height=3.5cm,width=.4\linewidth]{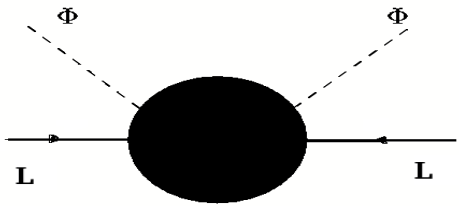}
    \caption{\label{fig:d-5} Dimension five operator responsible for
      neutrino mass~\cite{Weinberg:1980bf}.}
\end{figure}

Once the electroweak symmetry breaks through the nonzero vacuum
expectation value (vev) $\vev{\Phi}$, Majorana neutrino masses
$\propto \vev{\Phi}^2$ are induced, in contrast to the masses of the
charged fermions which arise from basic renormalizable interactions,
and are linear in $\vev{\Phi}$.
Moreover, the dimension-five operator $\O$ violates lepton number by
two units ($\Delta L=2$), whereas the charged fermion masses arise
from renormalizable L-conserving Yukawa interactions. This naturally
accounts for the smallness of neutrino masses irrespective of the
specific origin of neutrino mass. From such general point of view the
emergence of Dirac neutrinos would be a surprise, justified only in
the presence of an ``accidental'' lepton number symmetry.  For
example, neutrinos could naturally get very small Dirac masses via
mixing with a bulk fermion in models involving extra
dimensions~\cite{Dienes:1998sb,Arkani-Hamed:1998vp,Ioannisian:1999sw}.
Barring such very special circumstances, gauge theories give rise to
Majorana neutrinos.

Little more can be said from first principles about the {\sl
  mechanism} giving rise to the operator in Fig.~\ref{fig:d-5}, its
associated mass {\sl scale} or its {\sl flavour structure}.  For
example, the strength $\lambda$ of the operator $\O$ may be suppressed
by a large scale $M_X$ in the denominator (top-down) scenario, leading
to
$$ m_{\nu} = \lambda_0 \frac{\vev{\Phi}^2}{M_X}, $$
where $\lambda_0$ is some unknown dimensionless constant.
Gravity, which in a sense "belongs" to the SM, could induce the
dimension-five operator $\O$, providing the first example of a
top-down scenario with $M_X = M_P$, the Planck
scale~\cite{deGouvea:2000jp}. In this case the magnitude of the
resulting Majorana neutrino masses are too small.

Alternatively, the strength $\lambda$ of the operator $\O$ may be
suppressed by small parameters (e.g. scales, Yukawa couplings) in the
numerator and/or loop-factors (bottom-up scenario).
Both classes of scenarios are viable and have many natural
realizations. While models of the top-down type are closer to the idea
of unification, bottom-up schemes are closer to experimental test.

Models of neutrino mass may also be classified according to whether
or not additional neutral heavy states are present, in addition to the 
three isodoublet neutrinos. As an example, such leptons could be \321 
singlet ``right-handed'' neutrinos. 
In what follows we first consider top-down, then bottom-up scenarios.

\subsection{Seesaw and related models}
\label{sec:top-down-scenario}

\begin{figure}[ht] \centering
   \includegraphics[height=3cm,width=.38\linewidth]{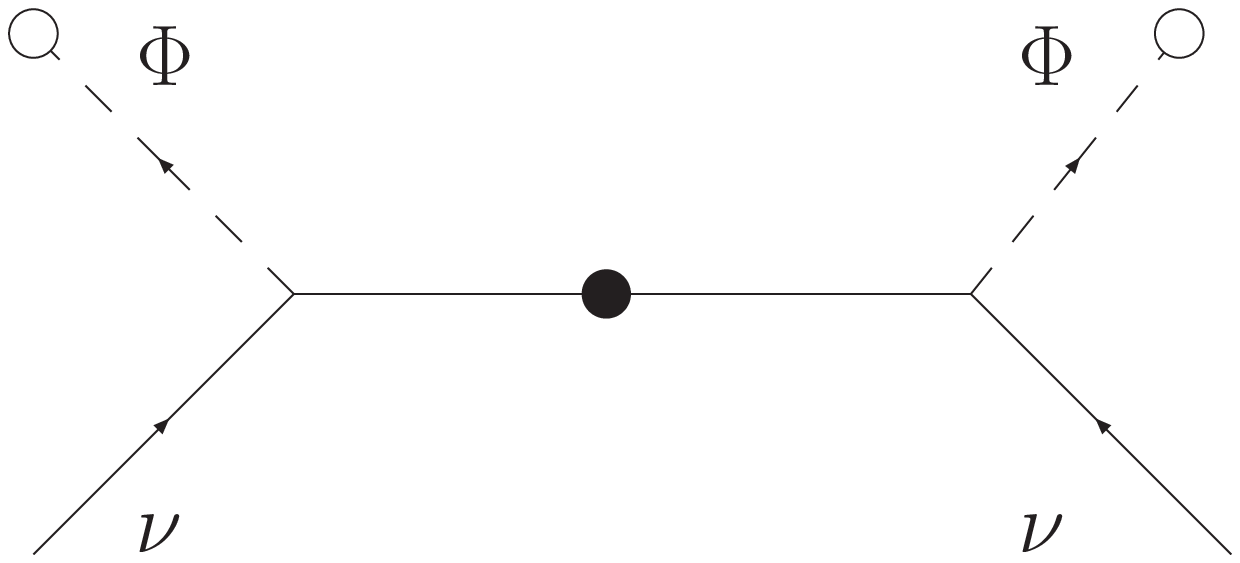} \hskip .8cm
    \includegraphics[height=3.7cm,width=.3\linewidth]{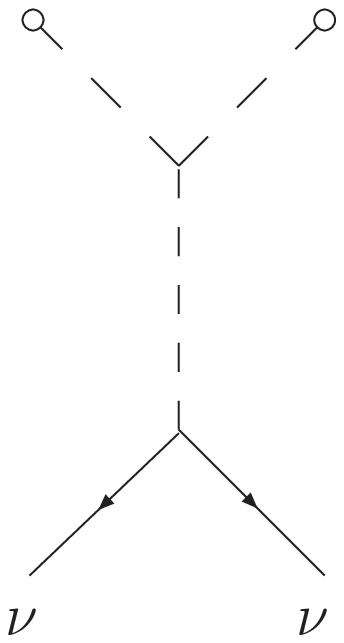}
    \caption{\label{fig:seesaw} %
      Two types of seesaw
      mechanism~\cite{Minkowski:1977sc,Orloff:2005nu,Weinberg:1980bf,schechter:1980gr,schechter:1982cv,Lazarides:1980nt}}
\end{figure}

The most popular top-down scenario is the
seesaw~\cite{Minkowski:1977sc}. The idea is to generate the operator
$\O$ by the exchange of heavy states, both fermions (type-I) and
scalars (type-II), as shown in Fig.~\ref{fig:seesaw}.
This can be implemented in many ways, with different gauge groups and
multiplet contents.  The main point is that, as the masses of the
intermediate states go to infinity, neutrinos naturally become
light~\cite{Orloff:2005nu}. The seesaw provides a simple realization
of Weinberg's operator~\cite{Weinberg:1980bf}.  Note that the seesaw
idea does not require the gauging of B-L, nor does it require it to be
broken spontaneously. In fact, most of the physics it encodes, and
which has been brilliantly confirmed by the recent oscillation
experiments, lies in its ``effective'' low-energy
form~\cite{schechter:1980gr}. Only in low-scale schemes like the
inverse-seesaw discussed in Sec. \ref{sec:inverse-seesaw} effects
associated with ``seesaw-dynamics'' may be observable (see
Secs.~\ref{sec:majoron} and \ref{sec:new-neutral-gauge}).

\subsubsection{``Effective''  seesaw}
\label{sec:effective-seesaw}

Much of the low energy phenomenology, such as that of neutrino
oscillations is blind to the details of the underlying seesaw theory
at high energies, e.~g. its gauge group, multiplet content or the
nature of B-L. For this purpose the most general way to describe the
physics of the seesaw is to characterize it, effectively, in terms of
the SM gauge structure~\cite{schechter:1980gr}. In the basis
$\nu_{L}$, $\nu^{c}_{L}$, corresponding to the three ``left'' and
three ``right'' neutrinos, respectively, the seesaw mass matrix has
SU(2) triplet, doublet and singlet terms described
as~\cite{schechter:1980gr}
\be
\label{ss-matrix-0} {\mathcal M_\nu} = \left(\begin{array}{cc}
    M_1 & D \\
    {D}^{T}   & M_2 \\ 
\end{array}\right) .
\ee 
Here we use the original notation of
reference~\cite{schechter:1980gr}, where the ``Dirac'' entry is
proportional to $\vev{\Phi}$, the $M_1$ comes from a triplet vev, and
$M_2$ may be added by hand, as it is a gauge singlet.  The particular
case $M_1=0$ was first mentioned in Ref.~\cite{Minkowski:1977sc}.

Note that, though symmetric, by the Pauli principle, the matrix
${\mathcal M_\nu}$ is complex, so that its Yukawa coupling
sub-matrices \(D\) as well as \(M_1\) and \(M_2\) are also complex
matrices, the last two symmetric. It is diagonalized by performing a
unitary transformation \(U_\nu\),
\begin{equation}
\label{eq:light-nu}
   \nu_i = \sum_{a=1}^{6}(U_\nu)_{ia} n_a ,
\end{equation}
so that
\begin{equation}
   U_\nu^T {\mathcal M_\nu} U_\nu = \mathrm{diag}(m_i,M_i) .
\end{equation}
This yields 6 mass eigenstates, including the three light neutrinos with
masses \(m_i\), and three two-component heavy leptons of masses \(M_i\).
The light neutrino mass states \(\nu_i\) are given in terms of the
flavour eigenstates via eq.~(\ref{eq:light-nu}).
The effective light neutrino mass, obtained this way is of the
form
\begin{equation}
  \label{eq:ss-formula0}
  m_{\nu} \simeq M_1 - D {M_2}^{-1} {D}^T . 
\end{equation}
The smallness of light neutrino masses is understood by assuming $M_2
\gg D \gg M_1$.  The above general structure forms the basis for the
description of the seesaw lepton mixing
matrix~\cite{schechter:1980gr}, given in
Sec.~\ref{sec:general-form-lepton}.

While it constitutes the most general description, and also the common
denominator of all seesaw schemes, such an ``effective'' seesaw does
not give a dynamical insight on the origin of neutrino mass.  For this
reason we now turn to schemes where lepton number symmetry is broken
spontaneously.

\subsubsection{The ``1-2-3'' seesaw mechanism}
\label{sec:majoron-seesaw}

The simplest possibility for the seesaw is to have ungauged lepton
number. It is also the most general, as it can be studied in the
framework of just the \321 gauge group. The mass terms in
eq.~(\ref{ss-matrix-0}) are given by triplet, doublet and singlet
vevs, respectively, as~\cite{schechter:1982cv}
\be
\label{ss-matrix-123} {\mathcal M_\nu} = \left(\begin{array}{cc}
    Y_3 v_3 & Y_\nu \vev{\Phi} \\
    {Y_\nu}^{T} \vev{\Phi}  & Y_1 v_1 \\
\end{array}\right) 
\ee 
As already mentioned, the Yukawa coupling sub-matrices \(Y_\nu\) as
well as \(Y_3\) and \(Y_1\) are complex matrices, the last two
symmetric.

The new dynamical insight provided by such ``1-2-3'' seesaw containing
singlet, doublet and triplet scalar multiplets, is that they obey a
simple vev seesaw relation of the type
 \begin{equation}
   v_3 v_1 \sim {v_2}^2 \:\:\: \mathrm{with} \:\:\: v_1 \gg v_2 \gg v_3 
 \label{eq:123-vev-seesaw}
 \end{equation}
 where $v_2 \equiv \vev{\Phi}$ denotes the SM Higgs doublet vev, fixed
 by the W-boson mass.  This hierarchy implies that the triplet vev
 $v_3 \to 0$ as the singlet vev $v_1$ grows. This is consistent with
 the minimization of the corresponding \321 invariant scalar
 potential, and implies that the triplet vev is ``induced''.

 Small neutrino masses arise either by heavy \321 singlet
 ``right-handed'' neutrino exchange (type I) or by the small effective
 triplet vev (type II), as illustrated in Fig.~\ref{fig:seesaw}.
 The effective light neutrino mass becomes,
\begin{equation}
  \label{eq:ss-formula-123}
  m_{\nu} \simeq Y_3 v_3 -
Y_\nu {Y_1}^{-1} {Y_\nu}^T \frac{{\vev{\Phi}}^2}{v_1}
\end{equation}
The corresponding seesaw diagonalization matrices can be given
explicitly as a systematic matrix perturbation series expansion in $D
M_2^{-1}$, given in Ref.~\cite{schechter:1982cv}.

In such ``1-2-3'' seesaw, since lepton number is ungauged, there is a
physical Goldstone boson associated with its spontaneous breakdown,
the majoron~\cite{chikashige:1981ui}. Its profile can be determined on
symmetry grounds and its couplings to neutrinos can be found
systematically, see~\cite{schechter:1982cv} for details.

\subsubsection{Left-right symmetric and SO(10)}
 \label{sec:left-right-symmetric}

 A more symmetric setting for the seesaw is a gauge theory containing
 B-L as a generator, such as \lr or the unified models based on \10 or
 \e6~\cite{Minkowski:1977sc,Orloff:2005nu,Lazarides:1980nt}.
 For example in \10 each matter generation is naturally assigned to a
 {\bf 16} (spinorial in \10) so that the {\bf 16} . {\bf 16} .  {\bf
   10} and {\bf 16} . {\bf 126} . {\bf 16} couplings generate all
 entries of the seesaw neutrino mass matrix,
\be
\label{ss-matrix} {\mathcal M_\nu} = \left(\begin{array}{cc}
    Y_L \vev{\Delta_L} & Y_\nu \vev{\Phi} \\
    {Y_\nu}^{T} \vev{\Phi}  & Y_{R} \vev{\Delta_R} \\
\end{array}\right)\,.
\ee 
Here the basis is $\nu_{L}$, $\nu^{c}_{L}$, as before, \(Y_L\) and
\(Y_R\) denote the Yukawas of the {\bf 126} of \10, whose vevs
$\vev{\Delta_{L,R}}$ give rise to the Majorana terms. They correspond
to \(Y_1\) and \(Y_3\) of the simplest ``1-2-3'' model.  On the other
hand ${Y_\nu}$ denotes the {\bf 16} . {\bf 16} .  {\bf 10} Dirac
Yukawa coupling.
In \10 one has a discrete parity symmetry which implies \(Y_L=Y_R\)
and \(Y_\nu=Y_\nu^T\) as recently emphasized in
Ref.~\cite{Akhmedov:2005np}. Since this may get broken, we prefer to
keep, for generality, \(Y_L, Y_R\) as independent.

Small neutrino masses are induced either by heavy \321 singlet
``right-handed'' neutrino exchange (type I) or heavy scalar boson
exchange (type II) as illustrated in Fig.~\ref{fig:seesaw}.
The matrix \(\mathcal{M_\nu}\) is diagonalized by a unitary mixing
matrix \(U_\nu\) as before. The diagonalization matrices can be worked
out explicitly as a perturbation series, using the same method of
Ref.~\cite{schechter:1982cv}. 
This means that the explicit formulas for the \(6\times6\) unitary
diagonalizing matrix \(U\) given in Ref.~\cite{schechter:1982cv} also
hold in the left-right case, provided one takes into account that $v_1
\to \vev{\Delta_R}$ and $v_3 \to \vev{\Delta_L}$.

The effective light neutrino mass, obtained this way is of the
form
\begin{equation}
  \label{eq:ss-formula}
  m_{\nu} \approx Y_L \vev{\Delta_L} -
Y_\nu {Y_R}^{-1} {Y_\nu}^T \frac{{\vev{\Phi}}^2}{\vev{\Delta_R}}
\end{equation}
We have the new vev seesaw relation
 \begin{equation}
  \vev{\Delta_{L}} \vev{\Delta_{R}} \sim \vev{\Phi}^2
 \label{eq:vev-seesaw}
 \end{equation}
 which naturally follows from minimization of the left-right symmetric
 scalar potential, together with the vev hierarchy
 \begin{equation}
  \vev{\Delta_{L}} \ll  \vev{\Phi} \ll \vev{\Delta_{R}}
 \label{eq:vev-hierarchy}
 \end{equation}
 This implies, as before, that both type I and type II contributions
 vanish as $\vev{\Delta_{R}} \to \infty$. Notice that, strictly
 speaking, this version of left-symmetry is inconsistent with type-I
 seesaw~\cite{Orloff:2005nu} and requires the full form of the seesaw
 mass matrix~\cite{schechter:1980gr,schechter:1982cv}. There are,
 however, tripletless left-right seesaw variants where a similar
 vev-seesaw formula holds, see
 Refs.~\cite{Akhmedov:1995vm,Barr:2005ss,Fukuyama:2005gg,Malinsky:2005bi}
 and Sec.~\ref{sec:low-b-l}.

 If one can arrange for the breakdown of parity invariance to be
 spontaneous, then the smallness of neutrino masses gets correlated to
 the observed maximality of parity violation in low-energy weak
 interactions, as stressed by Mohapatra and
 Senjanovic~\cite{Orloff:2005nu}. However elegant this connection may
 be, it is phenomenologically not relevant, in view of the large value
 of the B-L scale. The latter is required both to fit the neutrino
 masses, as well as to unify the gauge couplings.
 Another important difference with respect to the simplest ``1-2-3''
 seesaw is the absence of the majoron, which is now absorbed as the
 longitudinal mode of the gauge boson corresponding to the B-L
 generator which picks up a huge mass.

 \subsubsection{``Double'' seesaw mechanism}
\label{sec:double-seesaw}

One can add any number of (anomaly-free) gauge singlet leptons $S_i$
to the SM, or any other gauge theory~\cite{schechter:1980gr}. For
example, in \10 and \e6 one may add leptons outside the {\bf 16} or
the {\bf 27}, respectively.  New important features may emerge when
the seesaw is realized with non-minimal lepton content.  Here we
mention the seesaw scheme suggested in Ref.~\cite{mohapatra:1986bd}
motivated by string theories~\cite{Witten:1985xc}.
The model extends minimally the particle content of the SM by the
sequential addition of a pair of two-component \321 singlet leptons,
$\nu_i^c, S_i$, with \(i\) a generation index running over \(1,2,3\).
In the \(\nu,\nu^c,S\) basis, the \(9\times9\) neutral leptons
mass matrix $\mathcal{M_\nu}$ is given as
\begin{equation}
\label{eqn:doubleSeesaw}
{\mathcal M_\nu}=\left(
   \begin{array}{ccc}
      0   & Y_\nu^T \vev{\Phi} & 0   \\
      Y_\nu  \vev{\Phi} & 0     & M^T \\
      0   & M     & \mu
   \end{array}\right),
\end{equation}
in the basis $\nu_{L}$, $\nu^{c}_{L}$, $S_{L}$. Again \(Y_\nu\) is an
arbitrary \(3\times3\) complex Yukawa matrix, \(M\) and \(\mu\) are
SU(2) singlet complex mass matrices, \(\mu\) being symmetric.
Notice that it has zeros in the $\nu_{L}$-$\nu_{L}$ and
$\nu^{c}_{L}$-$\nu^{c}_{L}$ entries, a feature of several
string models~\cite{Witten:1985xc}. 

\begin{figure}[h] \centering
    \includegraphics[height=3cm,width=.45\linewidth]{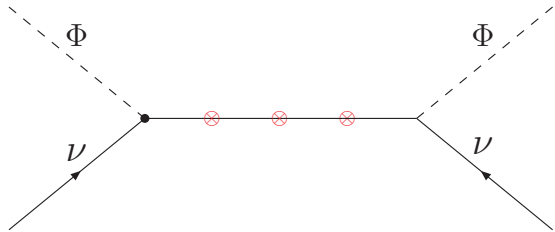}
    \caption{\label{fig:iss-mass} ``Double'' and ``inverse'' seesaw
      mechanism~\cite{mohapatra:1986bd}.}
\end{figure}

For $\mu \gg M$ one has to first approximation that the $S_i$ decouple
leaving the simpler seesaw at scales below that. In such a ``double''
seesaw scheme the three light neutrino masses are determined from
\begin{equation}
\label{eqn:lightNu}
    m_\nu =  {\vev{\Phi}^2} Y_\nu^T { M^{T}}^{-1} \mu M^{-1}  Y_\nu,
\end{equation}
The mass generation is illustrated in Fig.~\ref{fig:iss-mass}.
A new feature is that there are two independent scales, \(\mu\) and
\(M\), of which only \(\mu\) breaks the B-L symmetry.  Both scales can
be large but, as we will see in Sec.~\ref{sec:inverse-seesaw}, it is
natural for \(\mu\) to be small~\cite{mohapatra:1986bd}, instead of
large, \(\mu \ll M\), see Sec.~\ref{sec:inverse-seesaw}.
For the case \(\mu =M\) one formally recovers the usual seesaw form,
this is useful to present some results in simplified form, as used in
Fig.~\ref{fig:mueg2}.

Irrespective of what sets its scale, the entry \(\mu\) may be
proportional to the vev of an \321 singlet scalar, in which case the
model contains a singlet majoron, see Sec.~\ref{sec:majoron}.

 \subsubsection{``Novel''  \10 seesaw mechanism}
 \label{sec:low-b-l}

 The seesaw is a mechanism which allows for many possible
 realizations.  Schemes leading to the same pattern of neutrino masses
 may differ in many other respects. Correspondingly, there are many
 types of seesaw.  In addition to type
 I~\cite{Minkowski:1977sc,Orloff:2005nu} \cite{Lazarides:1980nt}, type
 II~\cite{schechter:1980gr,schechter:1982cv}, there are extended
 seesaw schemes, like
 type-III~\cite{Akhmedov:1995vm,Barr:2005ss,Fukuyama:2005gg} and the
 double/inverse seesaw described above.

\begin{figure}[h] \centering
    \includegraphics[height=3cm,width=.45\linewidth]{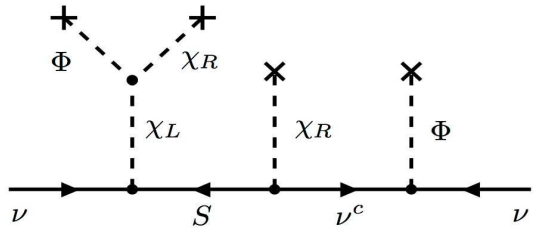}
    \caption{\label{fig:new-seesaw} Low B-L scale \10 seesaw
      mechanism~\cite{Malinsky:2005bi}.}
\end{figure}

 Here I turn to yet another seesaw that has recently been suggested in
 Ref.~\cite{Malinsky:2005bi}. It belongs to the class of
 supersymmetric \10 models with broken D-parity.  In addition to the
 states in the {\bf 16}, it contains three sequential gauge singlets
 $S_{iL}$ with the following mass matrix
\be \label{ess-matrix}
{\mathcal M_\nu} =
\left(\begin{array}{ccc}
0 & Y_\nu \vev{\Phi} & F \vev{\chi_L} \\
{Y_\nu}^{T} \vev{\Phi} & 0 & \tilde F \vev{\chi_R}     \\
F^{T} \vev{{\chi}_L}    & \tilde F^{T} \vev{\chi_R} & 0
\end{array}\right), 
\ee 
in the same basis $\nu_{L}$, $\nu^{c}_{L}$, $S_{L}$ as previously.
Notice that it has zeros along the diagonal, specially in the
$\nu_{L}$-$\nu_{L}$ and $\nu^{c}_{L}$-$\nu^{c}_{L}$ entries, thanks to
the fact that there is no {\bf 126}, a feature of several string
models~\cite{Witten:1985xc,mohapatra:1986bd}. The resulting light
neutrino mass matrix is
\begin{eqnarray}
m_{\nu} & \simeq &
\frac{\vev{\Phi}^2}{M_\mathrm{unif}} 
\left[Y_\nu ( F \tilde F^{-1})^{T}+( F \tilde F^{-1}) {Y_\nu}^{T}\right]
\end{eqnarray}
where $M_\mathrm{unif}$ is the unification scale, $F$ and $\tilde F$
denote independent combinations of Yukawa couplings of the $S_{iL}$.
One can see that the neutrino mass is suppressed by the unification
scale $M_{\mathrm{unif}}$ {\sl independently of the B-L breaking scale}.
In contrast to all previous seesaws, this one is {\sl linear} in the
Dirac Yukawa couplings $Y_\nu$, as illustrated in
Fig.~\ref{fig:new-seesaw}.
It is rather remarkable that one can indeed take the B-L scale as low
as TeV without generating inconsistencies, neither with neutrino
masses, nor with gauge coupling unification~\cite{Malinsky:2005bi}.

\subsection{Low-scale models}
\label{sec:bottom-up-scenario}

There are many models of neutrino mass where the operator $\O$ is
induced from physics at accessible scales, TeV or less.  The smallness
of its strength may be naturally achieved due to loop and Yukawa
couplings suppression. Moreover, the strength of the operator $\O$ may
be suppressed by small lepton number violating parameters that appear
in its numerator, instead of its denominator, as commonly assumed. The
latter correspond to the seesaw-type schemes previously discussed.
The former correspond to the class we are about to describe. Before I
do so, let me emphasize that such models are also ``natural'' in
t'Hooft's sense~\cite{'tHooft:1979bh}: {\sl ``an otherwise arbitrary
  parameter may be taken as small when the symmetry of the Lagrangean
  increases by having it vanish''}.

Since all particles are at the TeV scale, these models naturally lead
to possibly testable phenomenological implications, including \lfv and
modifications in muon and tau decays.

Moreover, when the breaking of lepton number entailed in these models
takes place spontaneously, the corresponding Goldstone boson has a
remarkable property: it may couple substantially to the SM Higgs
boson, which can therefore have a sizeable invisible decay branching
ratio~\cite{Joshipura:1993hp,romao:1992zx,Hirsch:2004rw,Hirsch:2005wd}
\begin{equation}
  \label{eq:JJ}
  H \to JJ
\end{equation}
where $J$ is the majoron.  This show that, although neutrino masses
are small, the neutrino mass generation may have very important
implications for the mechanism of electroweak symmetry breaking.
One must therefore take into account the existence of the invisible
channel in designing Higgs boson search strategies at future collider
experiments~\cite{deCampos:1997bg,Abdallah:2003ry}. Further discussion
in Sec.~\ref{sec:majoron}.

\subsubsection{Radiative models}
\label{sec:radiative-models}

\begin{figure}[h] \centering
    \includegraphics[height=3cm,width=.45\linewidth]{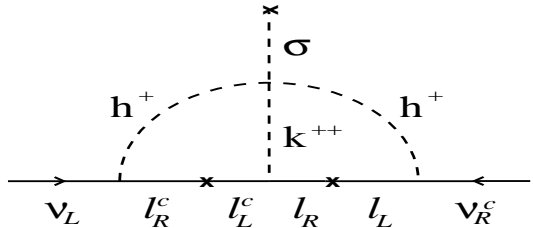}
    \caption{\label{fig:neumass} 
    Two-loop origin for neutrino mass~\cite{babu:1988ki,Peltoniemi:1993pd}.}
\end{figure}

Neutrino masses may be induced by calculable radiative
corrections~\cite{zee:1980ai}. For example, they can arise at the
two-loop level~\cite{babu:1988ki} as illustrated in
Fig.~\ref{fig:neumass}. Up to a logarithmic factor one has,
schematically,
\begin{equation}
   \label{eq:babu}
{\mathcal M_\nu} \sim \lambda_0 \left(\frac{1}{16\pi^2}\right)^2 
f Y_l h Y_l f^T \frac{\vev{\Phi}^2}{(m_k)^2} \vev{\sigma}
 \end{equation}
 in the limit where the doubly-charged scalar $k$ is much heavier than
 the singly charged one. Here $l$ denotes a charged lepton, $f$ and
 $h$ are their Yukawa coupling matrices and $Y_l$ denotes the SM Higgs
 Yukawa couplings to charged leptons. Here $\vev{\sigma}$ denotes an
 \321 singlet vev used in Ref.~\cite{Peltoniemi:1993pd}.  Clearly,
 even if the proportionality factor $\lambda_0$ is large, the neutrino
 mass can be made naturally small by the presence of a product of five
 small Yukawas and the appearance of the two-loop factor.

\subsubsection{Supersymmetry and neutrino mass}
\label{sec:supersymm-as-orig}

Low energy supersymmetry can be the origin of neutrino
mass~\cite{Hirsch:2004he}.
The intrinsically supersymmetric way to break lepton number is to
break the so-called R parity. This could happen spontaneously, driven
by a nonzero vev of an \321 singlet
sneutrino~\cite{Masiero:1990uj,romao:1992vu,romao:1997xf}, and leads
to an effective model characterized by purely bilinear R parity
violation~\cite{Diaz:1998xc}. This also serves as reference model, as
it provides the minimal way to add neutrino masses to the MSSM, we
call it RMSSM.
Neutrino mass generation takes place in a hybrid scenario, with one
scale generated at tree level and the other induced by ``calculable''
radiative corrections~\cite{Hirsch:2000ef}.
\begin{figure}[h] \centering
    \includegraphics[height=3.2cm,width=.5\linewidth]{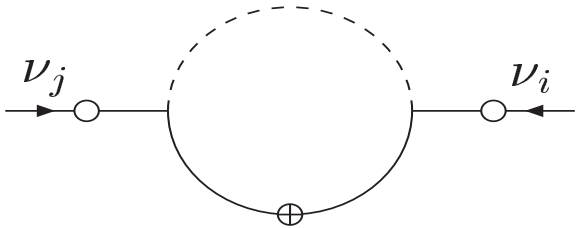}
    \caption{\label{fig:rpsusy} Loop origin of solar mass scale in
      RMSSM~\cite{Hirsch:2000ef}.}
\end{figure}
Here the two blobs in the graph Fig.~\ref{fig:rpsusy} denote $\Delta
L=1$ insertions, while the crossed blob accounts for chirality
flipping, and $A$ denotes the trilinear soft supersymmetry breaking
coupling. The general form of the expression is quite involved but the
approximation
\begin{equation}
   \label{eq:bilinear}
{\mathcal M_\nu} \sim \left(\frac{1}{16\pi^2}\right) {\vev{\Phi}^2} \frac{A}{m_0} Y_d Y_d 
 \end{equation}
 holds in some regions of parameters.  The neutrino mass spectrum
 naturally follows a normal hierarchy, with the atmospheric scale
 generated at the tree level and the solar mass scale arising from
 calculable loops.

\subsubsection{``Inverse'' seesaw mechanism}
\label{sec:inverse-seesaw}

Here we mention that there are also low-scale tree-level neutrino mass
schemes with naturally light neutrinos. One is the inverse seesaw
scheme suggested in \cite{mohapatra:1986bd} and already described
in Sec.~\ref{sec:double-seesaw}.
The mass matrix is the same as that of the double seesaw model, given
in eq.~(\ref{eqn:doubleSeesaw}). The only difference is that now the
entry \(\mu\) is taken very small, e.~g.  \(\mu \ll Y_\nu \vev{\phi}
\ll M\)~\cite{mohapatra:1986bd}.  Notice that for small \(\mu\)
neutrino masses vanish with \(\mu\),
$$m_\nu =  {\vev{\Phi}^2} Y_\nu^T { M^{T}}^{-1} \mu M^{-1}  Y_\nu $$
as illustrated in Fig.~\ref{fig:iss-mass}.

The fact that the neutrino mass vanishes as \(\mu\ \to 0\) is just the
opposite of the behaviour of the seesaw formulas in
Eqs.~(\ref{eq:ss-formula0}) (\ref{eq:ss-formula-123}) and
(\ref{eq:ss-formula}); thus this is sometimes called {\sl inverse}
seesaw model of neutrino masses.
The entry \(\mu\) may be proportional to the vev of an SU(2) singlet
scalar, in which case spontaneous B-L violation leads to the existence
of a majoron~\cite{gonzalez-garcia:1989rw}. This would be
characterized by a relatively low scale, so that the corresponding
phase transition could take place after the electroweak transition.
The naturalness of the model stems from the fact that in the limit
when $\mu \to 0$ lepton number is recovered, increasing the symmetry
of the theory.

\section{The lepton mixing matrix}
\label{sec:lepton-mixing-matrix}

In any gauge theory in order to identify physical particles one must
diagonalize all relevant mass matrices, which typically result from
gauge symmetry breaking.
Mechanisms giving mass to neutrinos generally imply the need
for new interactions whose Yukawas (like $Y_\nu$ in seesaw-type
schemes) will coexist with that of the charged leptons, $Y_l$.
Since in general these are independent one has that, like quarks,
massive neutrinos generally mix. The structure of this mixing is not
generally predicted from first principles.
Whatever the ultimate high energy gauge theory may be it must be
broken to the SM at low scales, so one should characterize the
structure of the lepton mixing matrix in terms of the \321 structure.
The procedure is the familiar one from the quark sector.

\subsection{Dirac case}
\label{sec:dirac-case}

Here we derive the structure of the lepton mixing matrix of massive
Dirac neutrinos and its parametrization, as presented in
Ref.~\cite{schechter:1980gr}. From the start $V$ can be parametrized
as
\begin{equation}
  \label{eq:par0}
V = \omega_0 (\gamma) \prod_{i<j}^{n} \omega_{ij} (\eta_{ij})\:
\end{equation}
where $$\omega_0 (\gamma)=\mathrm{exp}\: i(\sum_{a=1}^{n} \gamma_a
A_a^a)$$ is a diagonal unitary ``Cartan'' matrix, described by $n-1$
real parameters $\gamma_a$~\footnote{By choosing an overall relative
  phase between charged leptons and Dirac neutrinos we can take $V$ as
  unimodular, i.~e.~det $V=1$, so that $\sum_{a=1}^{n} \gamma_a =1$.}.
On the other hand each factor
$$\omega_{ab} (\eta_{ab})=\mathrm{exp}\: \sum_{a=1}^{n} (\eta_{ab}A_a^b - \eta_{ab}^*
A_b^a)$$ is a complex rotation in ${ab}$ with parameter $\eta_{ab} =
|\eta_{ab}| \mathrm{exp}\:i\theta_{ab}$.  For example,
\begin{equation}
  \label{eq:par12}
\omega_{12} (\eta_{12}) =
\left(\begin{array}{ccccc}
c_{12} & e^{i \phi_{12}} s_{12} & 0 ~~~...\\
-e^{-i \phi_{12}} s_{12} & c_{12} & 0 ~~~...\\
0  & 0 & 1 ~~~...\\
...&....&...~~~...
\end{array}\right)
\end{equation}
Once the charged leptons and Dirac neutrino mass matrices are
diagonal, one can still rephase the corresponding fields by $\omega_0
(\alpha)$ and $\omega_0 (\gamma-\alpha)$, respectively, keeping
invariant the form of the free Lagrangean. This results in the form
\begin{equation}
  \label{eq:par}
V = \omega_0 (\alpha) \prod_{i<j}^{n} \omega_{ij} (\eta_{ij})\:
\omega_0^\dagger (\alpha)\,,  
\end{equation}
where we are still free to choose the $n-1$ $\alpha$-values associated
to Dirac neutrino phase redefinitions.  Using the conjugation property
\begin{equation}
  \label{eq:par-conj}
\omega_0 (\alpha) \omega_{ab} (|\eta_{ab}| \mathrm{exp}\:i\theta_{ab})\:
\omega_0^\dagger (\alpha)  = \omega_{ab} [|\eta_{ab}| \mathrm{exp}\:i(\alpha_a + \theta_{ab} - \alpha_b )]
\end{equation}
we arrive at the final Dirac lepton mixing matrix which is, of course,
identical in form to that describing quark mixing.
It involves a set of 
\begin{equation}
  \label{eq:count}
n(n-1)/2 \:  \: \:\mathrm{mixing  \: \: angles} \: \: \theta_{ij} \: \: \mathrm{and} \: \: \: n(n-1)/2-(n-1) 
\mathrm{ \: CP  \: \:phases}\,.   
\end{equation}
where $(n-1)$ phases were eliminated.  This is the parametrization as
originally given in Ref.~\cite{schechter:1980gr}, with unspecified
factor ordering.
From Eq.~(\ref{eq:count}) one sees that for $n=3$ there are 3 angles
and precisely one leptonic CP violating phase, just as in the
Kobayashi-Maskawa matrix describing quark mixing.
Two of the three angles are involved in solar and atmospheric
oscillations, so we set $\theta_{12} \equiv \theta_\Sol$ and
$\theta_{23} \equiv \theta_\Atm$.  The last angle in the
three--neutrino leptonic mixing matrix is $\theta_{13}$,
\begin{equation}
   \label{eq:w13}
\omega_{13} = \left(\begin{array}{ccccc}
c_{13} & 0 & e^{i \Phi_{13}} s_{13} \\
0 & 1 & 0 \\
-e^{-i \Phi_{13}} s_{13} & 0 & c_{13}
\end{array}\right)\,.
 \end{equation}
 A convenient ordering prescription is to take {\bf 23 $\times$ 13
   $\times$ 12}, or ``atmospheric'' $\times$ ``reactor'' $\times$
 ``solar'', with the phase being associated to $\theta_{13}$.
 In summary, if neutrinos masses are added {\sl a la Dirac} their
 charged current weak interaction has exactly the same structure as
 that of quarks.

\subsection{Majorana case}
\label{sec:unit-appr-lept}

Here we consider the form of the lepton mixing matrix in models where
neutrino masses arise in the absence of right-handed neutrinos, such
as those in Sec.~\ref{sec:radiative-models}. The unitary form also
holds, to a good approximation, to models where SU(2) doublet
neutrinos mix only slightly with other states, like high-scale seesaw
models.

For $n$ generations of Majorana neutrinos the lepton mixing matrix has
exactly the same form given in Eq.~(\ref{eq:par}). The difference is
that in the case of Majorana neutrinos their mass term is manifestly
not invariant under rephasings of the neutrino fields. As a result,
the parameters $\alpha$ in Eq.~(\ref{eq:par}) can {\sl not} be used to
eliminate $n-1$ Majorana phases as we just did in
Sec.~\ref{sec:dirac-case}.  Consequently these are additional
phases~\cite{schechter:1980gr} which show up in L-violating
processes~\cite{Schechter:1981gk,doi:1981yb}. Such new sources of CP
violation in gauge theories with Majorana neutrinos are called
``Majorana phases''.  They already exist in a theory with just two
generations of Majorana neutrinos~\cite{schechter:1980gr}, $n=2$,
whose mixing matrix is described by
\begin{equation}
   \label{eq:w12}
\omega_{13} = \left(\begin{array}{ccccc}
c_{12} &  e^{i \Phi_{12}} s_{12} \\
-e^{-i \Phi_{12}} s_{12} &  c_{12}
\end{array}\right)\,,
 \end{equation}
 where $\phi_{12}$ is the Majorana phase (recall that Cabibbo mixing
 has no CP phase).  Such ``Majorana'' CP phases are, in a sense,
 mathematically more ``fundamental'' than the Dirac phase, whose
 existence, as we just saw, requires at least three generations.

 For the case of three neutrinos the lepton mixing matrix can be
 parametrized as~\cite{schechter:1980gr}
\begin{equation}
  \label{eq:2227}
K =  \omega_{23} \omega_{13} \omega_{12}
\end{equation}
where each factor in the product of the $\omega$'s is effectively
$2\times 2$, characterized by an angle and a CP phase. 
 Such symmetrical parameterization of the lepton mixing matrix, $K$
 can be written as:
\begin{equation}
\left[ \begin{array}{c c c}
c_{12}c_{13}&s_{12}c_{13}e^{i{\phi_{12}}}&s_{13}e^{i{\phi_{13}}}\\
-s_{12}c_{23}e^{-i{\phi_{12}}}-c_{12}s_{13}s_{23}e^{i({\phi_{23}}-{\phi_{13}})}
&c_{12}c_{23}-s_{12}s_{13}s_{23}
e^{i({\phi_{12}}+{\phi_{23}}-{\phi_{13}})}&c_{13}s_{23}e^{i{\phi_{23}}}\\
s_{12}s_{23}e^{-i({\phi_{12}}+{\phi_{23}})}-c_{12}s_{13}c_{23}e^{-i{\phi_{13}}}
&-c_{12}s_{23}e^{-i{\phi_{23}}}-
s_{12}s_{13}c_{23}e^{i({\phi_{12}}-{\phi_{13}})}&c_{13}c_{23}\\
\end{array} \right]. \nonumber
\label{writeout}
\end{equation}
All three CP phases are physical~\cite{Schechter:1981gk}:
$\phi_{12},\phi_{23}$ and $\phi_{13}$.  The ``invariant" combination
$\delta =\phi_{12} +\phi_{23}-\phi_{13}$ corresponds to the ``Dirac
phase". If neutrinos are of Dirac type, only a single phase (say
$\phi_{13}$) may be taken to be non-zero. This phase corresponds to
the phase present in the Kobayashi-Maskawa matrix, and this is the one
that affects neutrino oscillations. The other two phases are
associated to the Majorana nature of neutrinos and show up only in
lepton-number violating processes, like neutrinoless double beta
decay~\cite{Schechter:1981gk,doi:1981yb}.


An important subtlety arises regarding the conditions for CP
conservation in gauge theories of massive Majorana neutrinos.  Unlike
the case of Dirac fermions, where CP invariance implies that the
mixing matrix should be real, in the Majorana case the condition
is~\cite{schechter:1981hw}
$$ K^* = K \eta$$
where $\eta = \mathrm{diag}(+,+,..., , ,..)$ is the signature matrix
describing the relative signs of the neutrino mass eigenvalues that
follow from diagonalizing the relevant Majorana mass matrix, if one
chooses to use real diagonalizing matrices~\cite{Wolfenstein:1981rk}.
Consequently say, for $n=2$, both $\phi_{12} = \pi/2$ and $\phi_{12} =
0$ correspond to CP conservation, as emphasized by Wolfenstein.
These important signs determine the CP properties of the neutrinos and
play a crucial role in \znbb.

\subsection{Seesaw-type mixing}
\label{sec:general-form-lepton}

The most general theory of neutrino mass is effectively described in
SM terms by $(n,m)$, $n$ being the number of \321 isodoublets and $m
\neq 0$ the number of \321 isosinglet two-component leptons (the case
$m = 0$ was given above). Here we assume an arbitrary number of gauge
singlets, since they carry no anomaly.  The usual seesaw in
Secs.~\ref{sec:effective-seesaw}, \ref{sec:majoron-seesaw} and
\ref{sec:left-right-symmetric} has $m = n = 3$, while the extended
seesaw in Secs.~\ref{sec:double-seesaw}, \ref{sec:low-b-l} and
\ref{sec:inverse-seesaw} have $m = 2n = 6$.  Isosinglets have in
general a gauge and Lorentz invariant Majorana mass term.
The procedure holds in any scheme of Majorana neutrino masses where
isosinglet and isodoublet mass terms coexist~\cite{schechter:1980gr}.

The effective form of such a ``seesaw'' lepton mixing matrix has, in
addition to Majorana phases, many doublet-singlet mixing parameters,
in general complex~\cite{schechter:1980gr}. Its general structure is
substantially more complex than ``usual'', being described by a
rectangular matrix, called $K$.
As a result one finds that leptonic mixing as well as CP violation may
take place even in the massless neutrino
limit~\cite{branco:1989bn,rius:1990gk}.

The existence of these neutral heavy leptons could be inferred from
low energy weak decay processes, where the neutrinos that can be
kinematically produced are only the light ones. 
The mixing matrix describing the charged weak interactions of the
light (mass-eigenstate) neutrinos is effectively non-unitary, since
the coupling of a given light neutrino to the corresponding charged
lepton is decreased by a certain factor associated with the heavy
neutrino coupling.
There are constraints on the strength of the such mixing matrix
elements that follow from weak universality and low energy weak decay
measurements, as well as from LEP.

The full weak charged current mixing matrix $K$ of the general $(n,
m)$ models involves
\begin{equation}
  \label{eq:count-seesaw}
n(n+2m-1)/2~~\mathrm{mixing~~angles~~\theta_{ij}}
~~~~~~\mathrm{and}~~n(n+2m- 1)/2~~\mathrm{CP~~phases~~\phi_{ij}}\,.   
\end{equation}
For the explicit parametrization the reader is referred to the
original paper, Ref.~\cite{schechter:1980gr}. One sees that, for
example, the usual seesaw model [labeled $(3,3)$ in our language] is
characterized by 12 mixing angles and 12 CP phases (both Dirac and
Majorana-type)~\cite{schechter:1980gr}.

This number far exceeds the corresponding number of parameters
describing the charged current weak interaction of quarks. As already
mentioned, the reason is twofold: (i) neutrinos are Majorana
particles, their mass terms are not invariant under rephasings, and
(ii) the isodoublet neutrinos in general mix with the \321 singlets.
As a result, there are far more physical CP phases that may play a
role in  neutrino oscillations and/or leptogenesis (see below).

Another important feature which arises in any theory based on \321
where isosinglet and isodoublet lepton mass terms coexist is that the
leptonic neutral current is non-trivial~\cite{schechter:1980gr}: there
are non-diagonal couplings of the Z to the mass-eigenstate neutrinos.
They are expressed as a projective hermitian matrix 
$$P = K^\dagger K.$$
This contrasts with the neutral current couplings of mass-eigenstate
neutrinos in schemes where lepton number is conserved
(Sec.~\ref{sec:dirac-case}) or where no isosinglet leptons are
present, i.e., $m = 0$ (Sec.~\ref{sec:unit-appr-lept}). In both cases,
just as for quarks, the neutral current couplings are diagonal
(Glashow-Iliopoulos-Maiani mechanism).

Before we close, note that, in a scheme with $m < n$, $n - m$
neutrinos will remain massless, while $2m$ neutrinos will acquire
Majorana masses, $m$ light and $m$ heavy~\cite{schechter:1980gr}.  For
example, in a model with $n = 3$ and $m = 1$ one has one light and one
heavy Majorana neutrino, in addition to the two massless ones. In this
case clearly there will be less parameters than present in a model
with $m = n$.

\section{Phenomenology }
\label{sec:phenomenology}

Obviously the first phenomenological implication of neutrino mass
models is the phenomenon of neutrino oscillations, required to account
for the current solar and atmospheric neutrino data.
The interpretation of the data relies on good calculations of the
corresponding fluxes~\cite{Bahcall:2004fg,Honda:2004yz}, neutrino
cross sections and response functions, as well as on an accurate
description of neutrino propagation in the Sun and the Earth, taking
into account matter effects~\cite{mikheev:1985gs,wolfenstein:1978ue}.

\subsection{Status of neutrino oscillations}
\label{sec:stat-neutr-oscill}

Current neutrino oscillation data have no sensitivity to CP violation.
Thus we neglect all phases in the analysis and take, moreover, the
simplest unitary 3-dimensional form of the lepton mixing matrix in
Eq.~(\ref{eq:2227}) with the three phases set to zero.
In this approximation oscillations depend on the three mixing
parameters $\sin^2\theta_{12}, \sin^2\theta_{23}, \sin^2\theta_{13}$
and on the two mass-squared splittings $\Dms \equiv \Delta m^2_{21}
\equiv m^2_2 - m^2_1$ and $\Dma \equiv \Delta m^2_{31} \equiv m^2_3 -
m^2_1$ characterizing solar and atmospheric neutrinos.  The hierarchy
$\Dms \ll \Dma$ implies that one can set $\Dms = 0$, to a good
approximation, in the analysis of atmospheric and accelerator data.
Similarly, one can set $\Dma$ to infinity in the analysis of solar and
reactor data.

The world's neutrino oscillation data and their analysis, as of June
2006, are given in Ref.~\cite{Maltoni:2004ei} and will not be repeated
here. The new developments are: new Standard Solar
Model~\cite{Bahcall:2005va}, new SNO salt data~\cite{Aharmim:2005gt},
latest K2K~\cite{Ahn:2006zz} and MINOS~\cite{Tagg:2006sx} data. These
are briefly described in Appendix C of hep-ph/0405172 (v5). In what
follows we summarize the updated results of the analysis which takes
into account all these new data.
Apart from the ``positive'' data already mentioned, the analysis also
includes the constraints from ``negative" oscillation searches at
reactor experiments, CHOOZ and Palo Verde.
                                                                               
\begin{figure}[t] \centering
    \includegraphics[width=.95\linewidth,height=9cm]{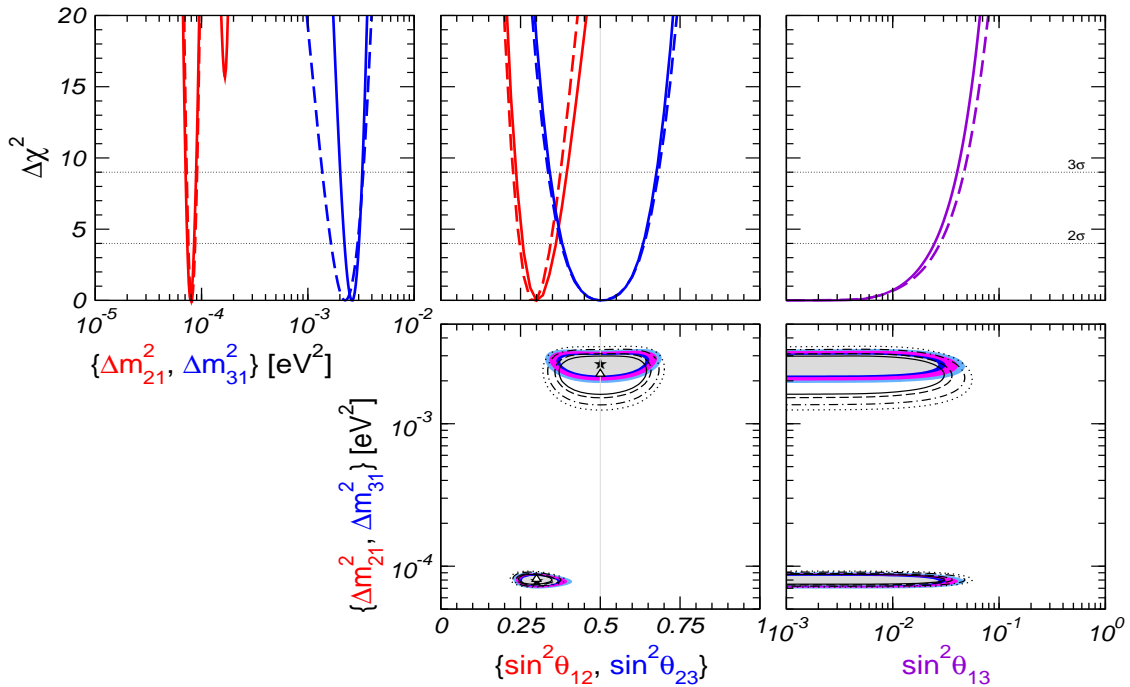}
    \caption{\label{fig:global} %
      Three--neutrino regions allowed by the world's neutrino
      oscillation data at 90\%, 95\%, 99\%, and 3$\sigma$ \CL\ for 2
      \dof\ as of June 2006, from Ref.~\cite{Maltoni:2004ei}. In top
      panels $\Delta \chi^2$ is minimized with respect to undisplayed
      parameters.}
\end{figure}

The three--neutrino oscillation parameters that follow from the global 
oscillation analysis in Ref.~\cite{Maltoni:2004ei} are summarized in 
Fig.~\ref{fig:global}.
In the upper panels of the figure the $\Delta \chi^2$ is shown as a
function of the three mixing parameters $\sin^2\theta_{12},
\sin^2\theta_{23}, \, \sin^2\theta_{13}$ and two mass squared
splittings $\Delta m^2_{21}, \Delta m^2_{31}$, minimized with respect
to the undisplayed parameters. The lower panels show two-dimensional
projections of the allowed regions in the five-dimensional parameter
space.  In addition to a confirmation of oscillations with $\Dma$,
accelerator neutrinos provide a better determination of $\Dma$. For
example, comparing dashed and solid lines in Fig.~\ref{fig:global} one
sees that the inclusion of the new data (mainly
MINOS~\cite{Tagg:2006sx}) leads to a slight increase in $\Dma$ and an
improvement on its determination (see \cite{Maltoni:2004ei} for
details).
On the other hand reactors~\cite{araki:2004mb} have played a crucial
role in selecting large-mixing-angle (LMA)
oscillations~\cite{pakvasa:2003zv} out of the previous ``zoo'' of
solutions~\cite{gonzalez-garcia:2000sq}.
The best fit values and the allowed 3$\sigma$ ranges of the
oscillation parameters from the global data are summarized in
Table~\ref{tab:summary}.
\begin{table}[t] \centering    \catcode`?=\active \def?{\hphantom{0}}
      \begin{tabular}{|l|c|c|}        \hline        parameter & best
      fit & 3$\sigma$ range         \\  \hline\hline        $\Delta
      m^2_{21}\: [10^{-5}~\eVq]$        & 7.9?? & 7.1--8.9 \\
      $\Delta m^2_{31}\: [10^{-3}~\eVq]$ & 2.6?? &  2.0--3.2 \\
      $\sin^2\theta_{12}$        & 0.30? & 0.24--0.40 \\
      $\sin^2\theta_{23}$        & 0.50? & 0.34--0.68 \\
      $\sin^2\theta_{13}$        & 0.00 & $\leq$ 0.040 \\
      \hline
\end{tabular}    \vspace{2mm}
\caption{\label{tab:summary} Neutrino oscillation parameters as of June 2006, 
from Ref.~\cite{Maltoni:2004ei}.}
\end{table}

Note that in a three--neutrino scheme CP violation disappears when two
neutrinos become degenerate or when one of the angles
vanishes~\cite{schechter:1980bn}.  As a result CP violation is doubly
suppressed, first by $\alpha \equiv \Dms/\Dma$ and also by the small
mixing angle $\theta_{13}$.
\begin{figure}[t]
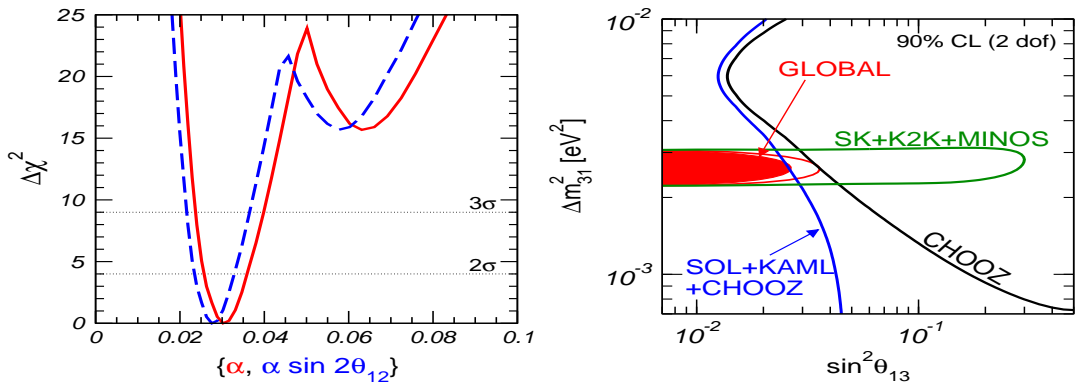
 \centering
  \includegraphics[height=5cm,width=.45\linewidth]{F-fcn.alpha06.eps}
\includegraphics[height=5cm,width=.45\linewidth]{th13-06.eps}
\caption{\label{fig:alpha}%
  Determination of $\alpha \equiv \Dms / \Dma$ and bound on
  $\sin^2\theta_{13}$ from data, as of June 2006, from
  Ref.~\cite{Maltoni:2004ei}.}
\end{figure}
The left panel in Fig.~\ref{fig:alpha} gives the parameter $\alpha$,
as determined from the global $\chi^2$ analysis.
The right panel shows the impact of different data samples on
constraining $\theta_{13}$.  One sees that, although for larger $\Dma$
values the bound on $\sin^2\theta_{13}$ is dominated by CHOOZ, this
bound deteriorates quickly as $\Dma$ decreases (see
Fig.~\ref{fig:alpha}), so that the solar and KamLAND data become
relevant.

There is now a strong ongoing effort aimed at probing $\theta_{13}$
and CP violation in future neutrino oscillation searches at reactors
and
accelerators~\cite{Alsharoa:2002wu,apollonio:2002en,albright:2000xi}.
As we saw, the basic parameters $\alpha \equiv \Dms/\Dma$ and
$\sin^2\theta_{13}$ characterizing the strength of CP violation in
neutrino oscillations are small.
Prospects for probing $\sin^2\theta_{13}$ at long baseline reactor and
accelerator neutrino oscillation experiments are given in
Ref.~\cite{Huber:2004ug},

Information on $\sin^2\theta_{13}$ may also come from a totally
different class of studies of the day/night effect in large water
Cerenkov solar neutrino experiments such as UNO or
Hyper-K~\cite{SKatm04}~\cite{Akhmedov:2004rq}.

\subsection{Predicting  neutrino masses and mixing}
\label{sec:pred-neutr-mixing}

Gauge symmetry alone is not sufficient to predict particle mixings,
neither for the quarks, nor for the leptons: such ``flavour problem''
has remained with us for a while.

As we saw in Sec.~\ref{sec:stat-neutr-oscill} five of the basic
parameters of the lepton sector are currently probed in neutrino
oscillation studies.  These point towards a well defined pattern of
neutrino mixing angles, quite distinct from that of quarks.  Such
pattern is not easy to account for in the context of unified schemes
where quarks and leptons are connected.
The data seem to indicate an intriguing complementarity between quark
and lepton mixing
angles~\cite{Raidal:2004iw,minakata-2004-70,Ferrandis:2004vp,Dighe:2006zk}.


There has been a rush of papers attempting to understand the values of
the leptonic mixing angles from underlying symmetries at a fundamental
level.  For example the following form of the neutrino mixing angles
has been proposed~\cite{Harrison:2002kp}
\begin{align}
\label{eq:hps}
\tan^2\theta_{\Atm}&=\tan^2\theta_{23}^0=1\\ \nonumber
\sin^2\theta_{\textrm{Chooz}}&=\sin^2\theta_{13}^0=0\\
\tan^2\theta_{\Sol}&=\tan^2\theta_{12}^0=0.5 .\nonumber
\end{align}
Such Harrison-Perkins-Scott pattern~\cite{Harrison:2002er} could
result from some kind of flavour symmetry, valid at a very high energy
scale where the dimension-five neutrino mass operator arises.

One approach to predict neutrino masses and mixing angles was the idea
that neutrino masses arise from a common seed at some ``neutrino mass
unification'' scale $M_X$~\cite{chankowski:2000fp}, very similar the
merging of the SM gauge coupling constants at high energies due to
supersymmetry~\cite{amaldi:1991cn}.
However, in its simplest form this very simple theoretical ansatz is
now inconsistent (at least if CP is conserved) with the current
observed value of the solar mixing angle $\theta_{12}$ inferred from
current data.


A more satisfactory and fully viable alternative realization of the
``neutrino mass unification'' idea employs an $A_4$ flavour symmetry
introduced by Ernest Ma, in the context of a seesaw
scheme~\cite{babu:2002dz}. Starting from three-fold degeneracy of the
neutrino masses at a high energy scale, a viable low energy neutrino
mass matrix can indeed be obtained in agreement with neutrino data as
well as constraints on \lfv in $\mu$ and $\tau$ decays.
The model predicts maximal atmospheric angle and vanishing
$\theta_{13}$,
$$\theta_{23}=\pi/4~~~\rm{and}~~~\theta_{13}=0\:.$$ 
Although the solar angle $\theta_{12}$ is unpredicted, one
expects~\footnote{There have been realizations of the $A_4$ symmetry
  that also predict the solar angle, e.~g.
  Ref.~\cite{Hirsch:2005mc}.}
$$\theta_{12}=\O(1).$$ 
When CP is violated $\theta_{13}$ becomes arbitrary and the Dirac
phase is maximal~\cite{Grimus:2003yn}.

Within such flavour symmetric seesaw scheme one can show that the
lepton and slepton mixings are intimately related. The resulting
slepton spectrum must necessarily include at least one mass eigenstate
below 200 GeV, which can be produced at the LHC. The prediction for
the absolute Majorana neutrino mass scale $m_0 \geq 0.3$ eV ensures
that the model will be tested by future cosmological tests and
$\beta\beta_{0\nu}$ searches.  Rates for lepton flavour violating
processes $l_j \to \l_i + \gamma$ typically lie in the range of
sensitivity of coming experiments, with BR$(\mu \to e \gamma) \gsim
10^{-15}$ and BR$(\tau \to \mu \gamma) > 10^{-9}$.  

Finally, we mention that there have been attempts to realize the
Harrison-Perkins-Scott mixing pattern in Eq.~(\ref{eq:hps}) at some
high energy scale, and to correct its predictions by renormalization
group evolution~\cite{Altarelli:2005yp,Hirsch:2006je}. For a survey of
related attempts see Ref.~\cite{Altarelli:2004za}.

\subsection{Absolute scale of neutrino mass and \znbb}
\label{sec:neutr-double-beta}

Neutrino oscillations are blind to whether neutrinos are Dirac or
Majorana.  As we have seen, on general grounds, neutrinos are expected
to be Majorana~\cite{schechter:1980gr}.
Neutrinoless double beta decay and other \lnv processes, such as
neutrino transition electromagnetic
moments~\cite{schechter:1981hw,Wolfenstein:1981rk}
\cite{pal:1982rm,kayser:1982br} are able to probe the basic nature of
neutrinos.

\begin{figure}[h]
  \centering
\includegraphics[width=6cm,height=4cm]{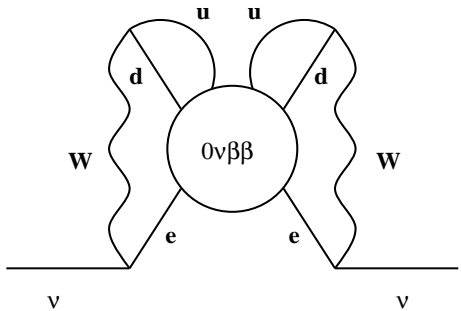}
\caption{Neutrinoless double beta decay and Majorana mass are
  equivalent~\cite{Schechter:1982bd}.}
 \label{fig:bbox}
\end{figure}

The significance of neutrinoless double beta decay stems from the
fact that, in a gauge theory, irrespective of the mechanism that
induces \znbb, it necessarily implies a Majorana neutrino
mass~\cite{Schechter:1982bd}, as illustrated in Fig.  \ref{fig:bbox}.
Thus it is a basic issue. Quantitative implications of the
``black-box'' argument are model-dependent, but the theorem itself
holds in any ``natural'' gauge theory.  


Conventional neutrino oscillations are also insensitive to the
absolute scale of neutrino
masses~\cite{bilenky:1980cx,Schechter:1981gk,doi:1981yb}.
Although the latter will be tested directly in high sensitivity
tritium beta decay studies~\cite{Drexlin:2005zt}, as well as by its
effect on the cosmic microwave background and the large scale
structure of the
Universe~\cite{Lesgourgues:2006nd,Hannestad:2006zg,Fogli:2004as} \znbb
may give valuable complementary information. For example, as seen
above, the $A_4$ model~\cite{babu:2002dz} gives a lower bound on the
absolute Majorana neutrino mass $m_{\nu}\gsim 0.3$ eV and may
therefore be tested in \znbb searches.

Now that oscillations are experimentally confirmed we know that \znbb
must be induced by the exchange of light Majorana neutrinos, the
so-called "mass-mechanism". The corresponding amplitude is
sensitive~\cite{Schechter:1981gk,doi:1981yb} both to the absolute
scale of neutrino mass, as well as to Majorana
phases~\cite{schechter:1980gr}, neither of which can be probed in
oscillations~\cite{bilenky:1980cx,Schechter:1981gk}.
\begin{figure}[h]
 \centering
 \includegraphics[clip,width=.7\linewidth,height=12cm]{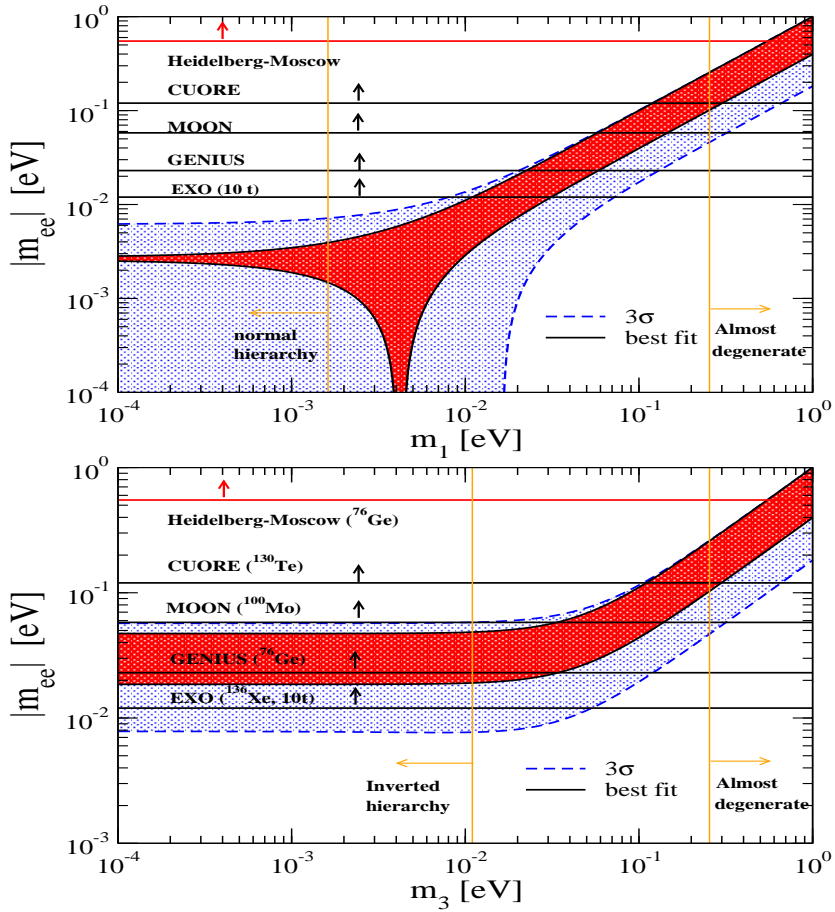}
 \caption{\znbb amplitude versus current oscillation data,
   from Ref.~\cite{Bilenky:2004wn}.}
\label{fig:nbbfut}
\end{figure}

Fig. \ref{fig:nbbfut} shows the estimated average mass parameter
characterizing the neutrino exchange contribution to \znbb versus the
lightest and heaviest neutrino masses.  The calculation takes into
account the current neutrino oscillation parameters in
\cite{Maltoni:2004ei} and state-of-the-art nuclear matrix
elements~\cite{Bilenky:2004wn}.
The upper (lower) panel corresponds to the cases of normal (inverted)
neutrino mass spectra. In these plots the ``diagonals'' correspond to
the case of quasi-degenerate
neutrinos~\cite{babu:2002dz}~\cite{caldwell:1993kn}~\cite{ioannisian:1994nx}
In the normal hierarchy case there is in general no lower bound on 
the \nbb rate since there can be a destructive interference amongst
the neutrino amplitudes. In contrast,  the inverted neutrino mass
hierarchy implies a ``lower'' bound for the \nbb amplitude.
\begin{figure}[h] \centering
    \includegraphics[width=.6\linewidth,height=5cm]{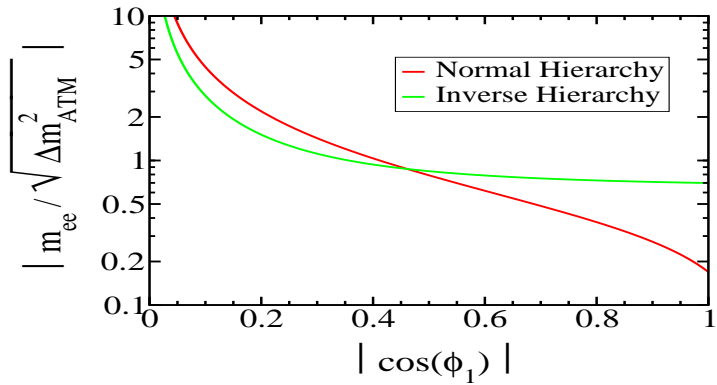}
    \caption{\label{fig:bbn-a4} %
      Lower bound on $|\vev{m_{ee}}|/\Dma$ vs $|\mathrm{cos}
      (\phi_1)|$ where $\phi_1$ is a Majorana phase in the model of
      Ref.~\cite{Hirsch:2005mc}. The lines in dark (red) and grey
      (green) correspond to normal and inverse hierarchy.}
\end{figure}
A specific normal hierarchy model for which a lower bound on \nbb can
be placed has been given in Ref.~\cite{Hirsch:2005mc}. An interesting
feature is that such lower bound depends, as expected, on the value of
the Majorana violating phase $\phi_1$, as indicated in
Fig.~\ref{fig:bbn-a4}.

The best current limit on $\meff$ comes from the Heidelberg-Moscow
experiment. There is also a claim made in
Ref.~\cite{Klapdor-Kleingrothaus:2004wj} (see also
Ref.~\cite{Aalseth:2002dt}) which will be important to confirm or
refute in future experiments. GERDA will provide an independent check
of this claim~\cite{Aalseth:2002rf}. SuperNEMO, CUORE, EXO, MAJORANA
and possibly other experiments will further extend the sensitivity of
current \nbb searches~\cite{dbd06}.

\subsection{Other phenomena}
\label{sec:other-phenomena}

If neutrino masses arise {\sl a la seesaw} the dynamics responsible
for generating the small neutrino masses seems most likely untestable.
In other words, beyond neutrino masses and oscillations, theories can
not be probed phenomenologically at low energies, due to the large
scale involved.
However when the breaking of lepton number symmetry is spontaneous
there is a dynamical ``tracer'' of the mass-generation mechanism
which might be probed experimentally. 

\subsubsection{Majoron physics}
\label{sec:majoron}

If neutrino masses follow from spontaneous violation of global lepton
number, the existence of the Goldstone boson brings new interactions
for neutrinos and Higgs boson(s) which may lead to new phenomena.
While neither of these is expected within the usual high-scale seesaw
schemes, both could lead to detectable signals in low-scale models of
neutrino mass.

As already mentioned, the majoron may couple substantially to the SM
Higgs boson, which can therefore have a sizeable decay branching
ratio~\cite{Joshipura:1993hp,romao:1992zx,Hirsch:2004rw,Hirsch:2005wd}
into the channel in eq.~(\ref{eq:JJ}). Such ``invisible'' channel is
experimentally detectable as missing energy or transverse momentum
associated to the Higgs~\cite{deCampos:1997bg,Abdallah:2003ry}.
Therefore in low-scale models of neutrino mass the neutrino
mass-giving mechanism may have a strong impact in the electroweak
sector.

Majoron-emitting neutrino decays can also be lead to detectable
signals in low-scale models of neutrino mass,
$$\nu_3 \to \nu + J,$$ 
For example, if the neutrinos decay in high density media, like
supernovae, characterized by huge matter densities, then the
``matter-assisted'' decays would lead to detectable signals in
underground water Cerenkov experiments~\cite{kachelriess:2000qc}.

\subsubsection{New  gauge boson}
\label{sec:new-neutral-gauge}

Although in the usual large-scale seesaw with gauged B-L symmetry,
there are new gauge bosons associated to the neutrino mass generation
these are too heavy to give any detectable effect.
Only when the B-L scale is low, as in the model discussed in
Sec.~\ref{sec:low-b-l} or the model considered in
Ref.~\cite{valle:1987sq}, there will exist a light new neutral gauge
boson, \ZP that could be detected in searches for Drell-Yan processes
at the LHC.

\subsubsection{Lepton flavour violation}
\label{sec:lept-flav-viol}

\begin{figure}[h] \centering
   \includegraphics[height=4cm,width=.8\linewidth]{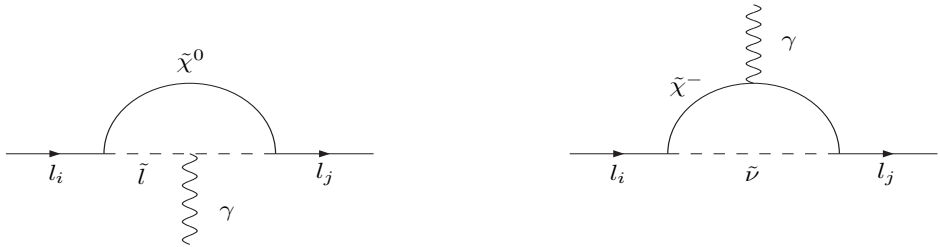}
\vglue -1cm
\caption{\label{fig:mueg1} Supersymmetric Feynman diagrams for
  \(l_{i}^{-} \to l_{j}^{-}\gamma\).}
\end{figure}

In the presence of supersymmetry, seesaw phenomenology is richer. A
generic feature of supersymmetric seesaw models is the existence of
processes with \lfv such as $\mu^-\to e^-\gamma$. Supersymmetry
contributes through the exchange of charginos (neutralinos) and
sneutrinos (charged
sleptons)~\cite{hall:1986dx,borzumati:1986qx,casas:2001sr}, as
illustrated in Fig.~\ref{fig:mueg1}.

\begin{figure}[h]
\centering
\includegraphics[clip,height=4.5cm,width=0.8\linewidth]{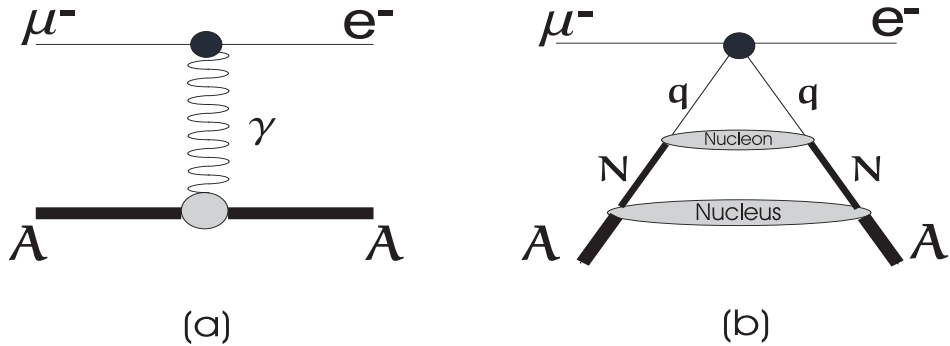}
\caption{Contributions to the nuclear $\mu^-- e^-$ conversion: (a)
  long-distance and (b) short-distance. For numerical results see
  Ref.~\cite{Deppisch:2005zm}}
     \label{fig:Diagrams}
\end{figure}

\begin{figure}[t] \centering
   \includegraphics[height=6cm,width=.45\linewidth]{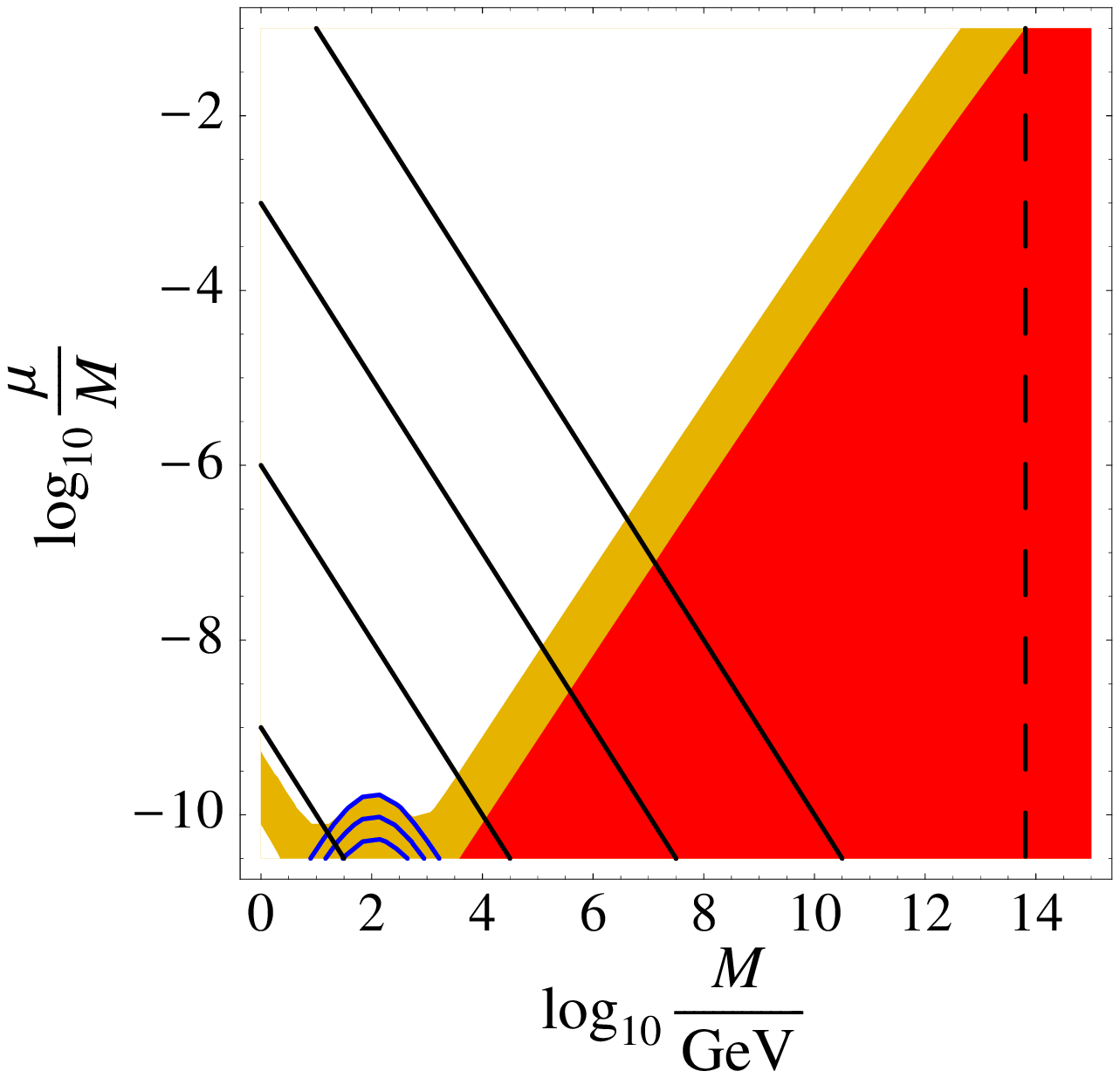}
   \includegraphics[height=5.5cm,width=.45\linewidth]{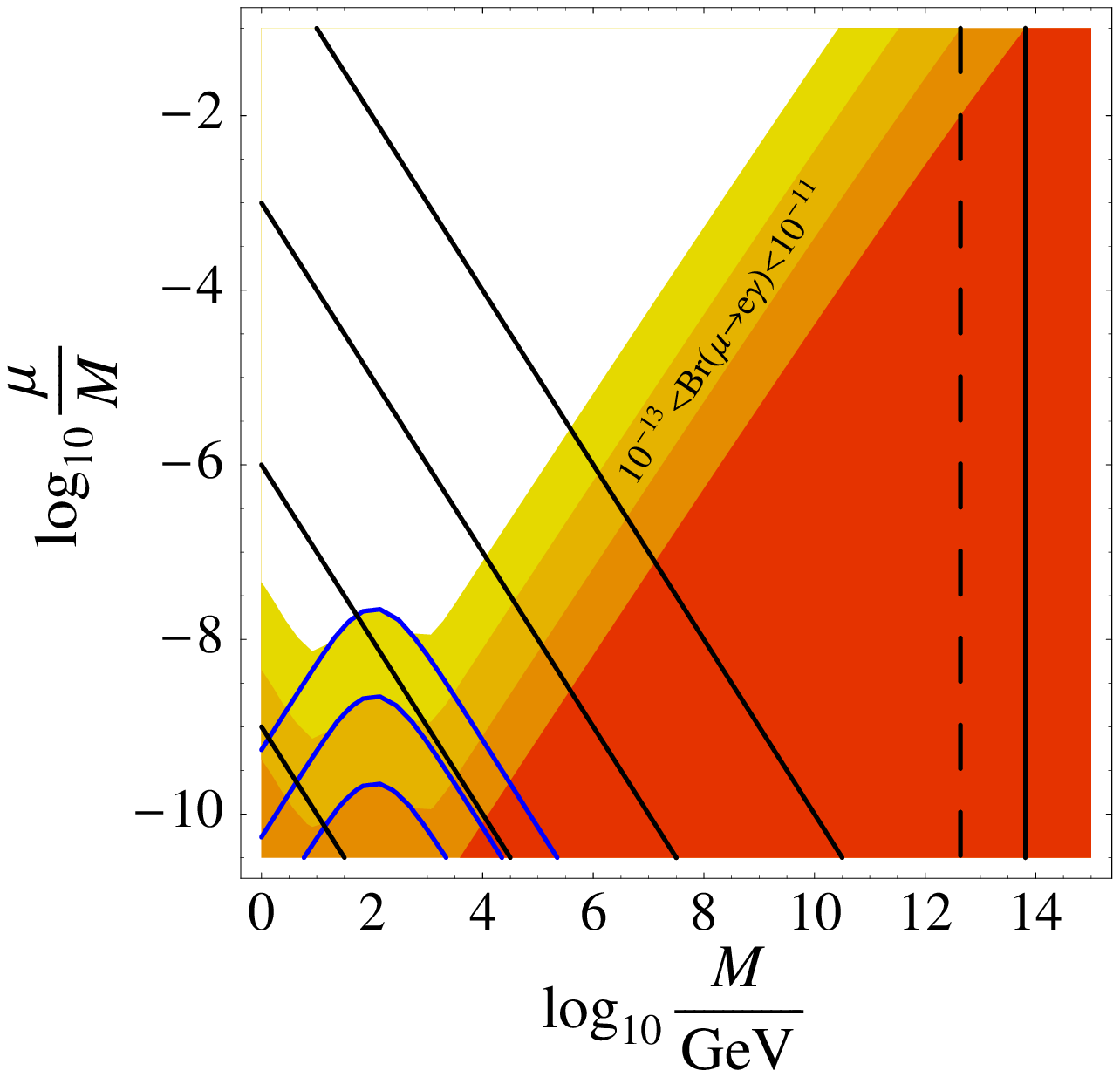}
   \caption{\label{fig:mueg2} \(Br(\mu\to e\gamma)\) in the
     supersymmetric inverse seesaw model of Ref.~\cite{Deppisch:2004fa}.}
\end{figure}

Similarly the nuclear $\mu^-- e^-$ conversion arises, as
indicated in Fig.~\ref{fig:Diagrams}.
The rates for these process can both be sizeable. As an example,
consider the rates for the $\mu^-\to e^-\gamma$ decay, given
in Fig.~\ref{fig:mueg2}.

The calculation leading to Fig.~\ref{fig:mueg2} is done in the
framework of the supersymmetric double/inverse seesaw model in
Secs.~\ref{sec:double-seesaw} and
\ref{sec:inverse-seesaw}~\cite{Deppisch:2004fa}.  This allows one to
analyse the interplay of neutral heavy
lepton~\cite{bernabeu:1987gr}~\footnote{Since in this model flavor and
  CP violation can occur in the massless neutrino limit, the allowed
  rates are unsuppressed by the smallness of neutrino
  masses~\cite{bernabeu:1987gr,branco:1989bn,rius:1990gk,gonzalez-garcia:1992be,Ilakovac:1994kj}.}
and supersymmetric contributions~\cite{casas:2001sr}.  The figure
shows the \(Br(\mu\to e\gamma)\) contours in the
\((M,\frac{\mu}{M})\)-plane (logarithmic scales) for hierarchical
light neutrinos with \(m_1=0\)~eV (left panel) and for degenerate
light neutrinos with \(m_1=0.3\)~eV (right panel). The shaded contours
correspond to (from left to right): \(Br(\mu\to e\gamma) =
10^{-15,-13,-11,-9}\).
For large \(M\) the estimates are similar to those of the standard
supersymmetric seesaw.
However, if the neutral heavy leptons are in the TeV range (a
situation not realizable in the minimal seesaw mechanism), the
\(Br(\mu\to e\gamma)\) rate can be enhanced even in the {\sl absence}
of supersymmetry.  This is indicated by the contours in the lower
left, which depict the contribution from neutral heavy leptons only.
For such \(M\) values around TeV or so, the quasi-Dirac neutral heavy
leptons may be directly produced at
accelerators~\cite{Dittmar:1990yg}.
Note that in order to have low \(M\) then \(\mu\) should be rather
small, to keep neutrinos light. This is indicated by the diagonal
lines, which indicate contours (top to bottom) of constant \(\mu=
1,10^{-3},10^{-6},10^{-9}\)~GeV.  The vertical lines are contours of
\(Br(\mu\to e\gamma)\) in the standard supersymmetric seesaw.

Similarly, the rates for the nuclear $\mu^--e^-$ conversion in
Fig.~\ref{fig:Diagrams}~\cite{Deppisch:2005zm} fall within the
sensitivity of future experiments such as PRISM~\cite{Kuno:2000kd}.

\subsubsection{Reconstructing neutrino mixing at accelerators}
\label{sec:reconstr-neutr-mixin}

Low-scale models of neutrino mass, considered in
Sec.~\ref{sec:bottom-up-scenario}, offer the tantalizing possibility
of reconstructing neutrino mixing at high energy accelerators, like
the "Large Hadron Collider" (LHC) and the "International Linear
Collider" (ILC).

A remarkable example is provided by the models where supersymmetry is
the origin of neutrino mass~\cite{Hirsch:2004he}, considered in
Sec.~\ref{sec:supersymm-as-orig}.
A general feature of these models is that, unprotected by any
symmetry, the lightest supersymmetric particle (LSP) is unstable. In
order to reproduce the masses indicated by current neutrino
oscillation data, the LSP is expected to decay inside the
detector~\cite{Hirsch:2000ef}~\cite{deCampos:2005ri}.

\begin{figure}[h]
\centering
\includegraphics[clip,height=5cm,width=0.4\linewidth]{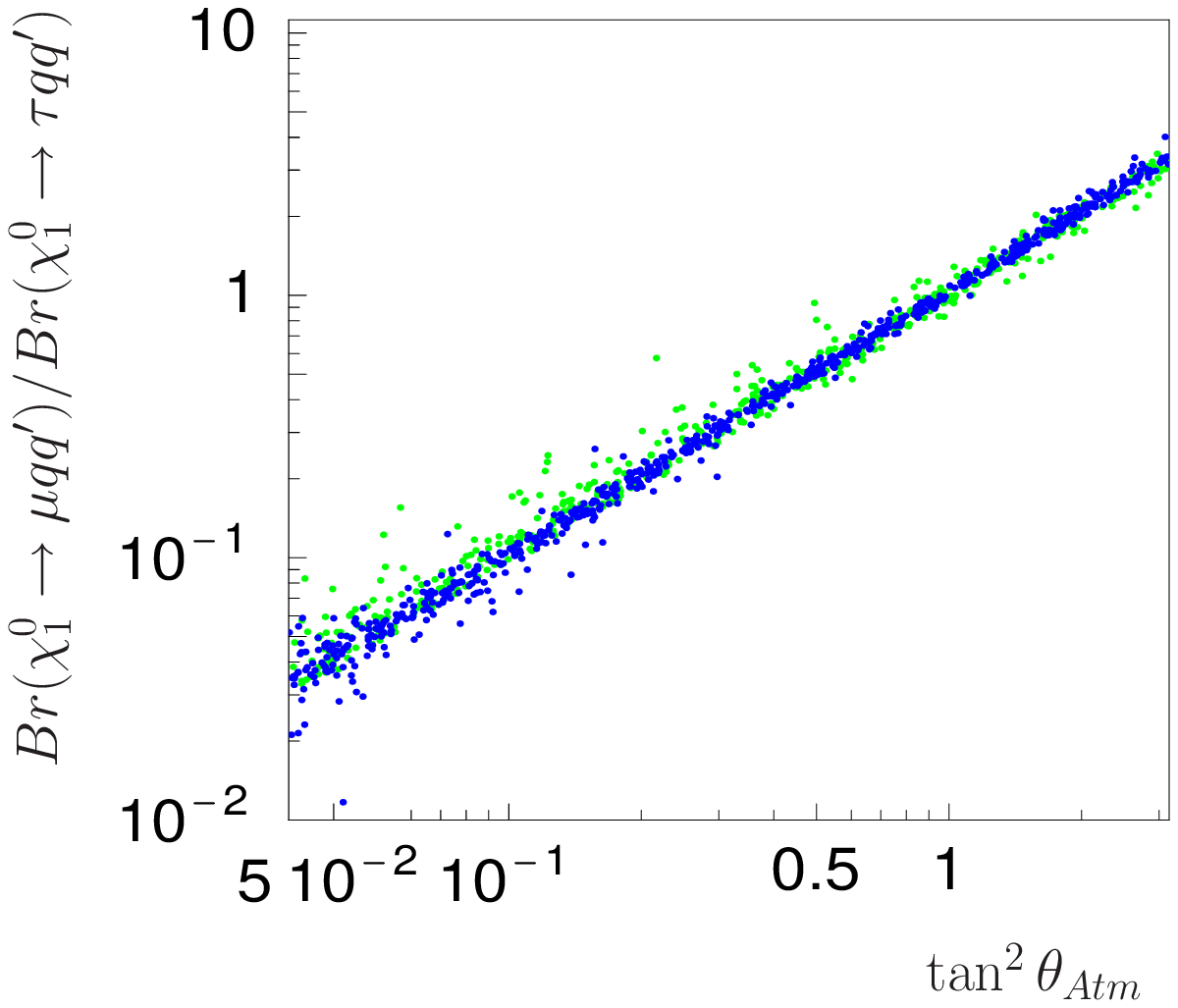}
\includegraphics[clip,height=5cm,width=0.4\linewidth]{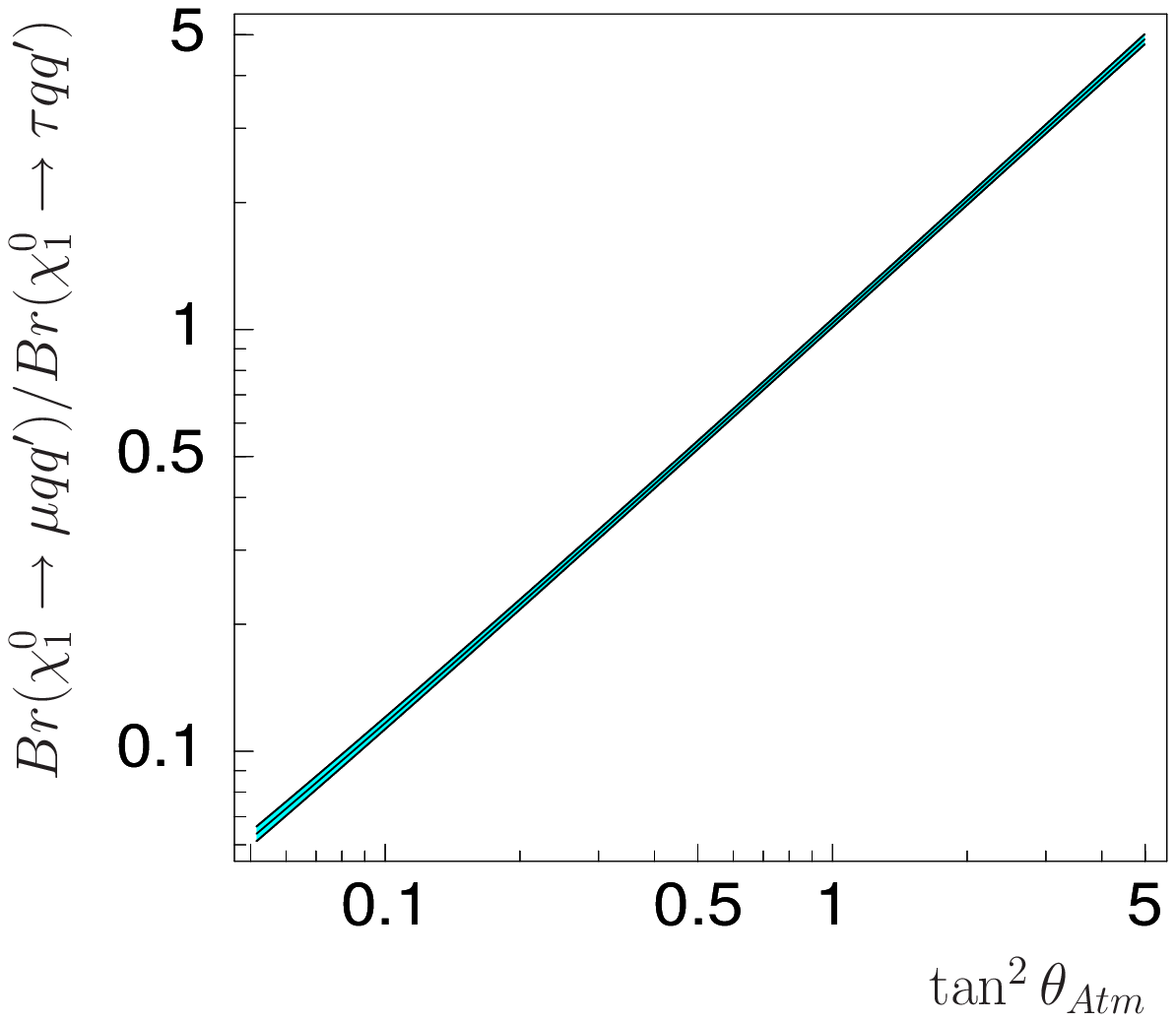}
\caption{LSP decays trace the atmospheric mixing
  angle~\cite{Porod:2000hv}.}
     \label{fig:correlation}
\end{figure}

More strikingly, LSP decay properties correlate with neutrino mixing
angles. For example, if the LSP is the lightest neutralino, it should
have the same decay rate into muons and taus, since the observed
atmospheric angle is close to
$\pi/4$~\cite{Porod:2000hv,romao:1999up,mukhopadhyaya:1998xj}.
Similar correlations hold irrespective of which supersymmetric
particle is the LSP~\cite{Hirsch:2003fe} and constitute a smoking gun
signature of this proposal that will be tested at upcoming
accelerators.

There are other examples of low-scale models that nicely illustrate
the possibility of probing neutrino properties at
accelerators~\cite{Ma:2000xh}.

\subsection{Thermal leptogenesis}
\label{sec:thermal-leptogenesis}

\begin{figure}[h]
\centering
\includegraphics[clip,height=3.2cm,width=0.8\linewidth]{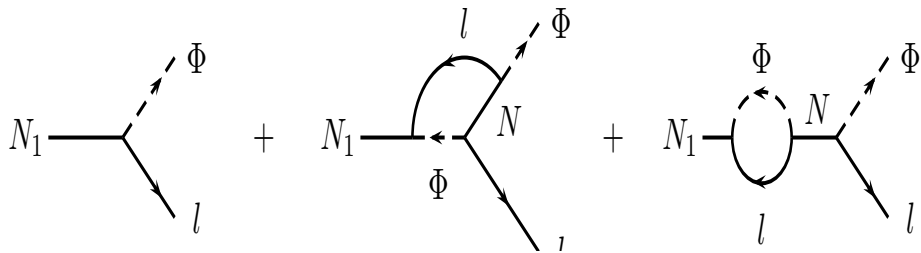}
\caption{Diagrams contributing to  leptogenesis.}
     \label{fig:lep-g}
\end{figure}

Now we briefly discuss one of the cosmological implications of
neutrino masses and mixing, in the context of seesaw schemes.  It has
long been noted~\cite{Fukugita:1986hr} that seesaw models open an
attractive possibility of accounting for the observed cosmological
matter-antimatter asymmetry in the Universe through the leptogenesis
mechanism~\cite{Buchmuller:2005eh}.
In this picture the decays of the heavy ``right-handed'' neutrinos
present in the seesaw play a crucial role.
These take place through diagrams in Fig.~\ref{fig:lep-g}. In order to
induce successful leptogenesis the decay must happen before the
electroweak phase transition~\cite{kuzmin:1985mm} and must also take
place out-of-equilibrium, i.~e. the decay rate must be less than the
Hubble expansion rate at that epoch. Another crucial ingredient is CP
violation in the lepton sector.
The lepton (or B-L) asymmetry thus produced then gets converted,
through sphaleron processes, into the observed baryon asymmetry.

In the framework of a supersymmetric seesaw scheme the high temperature
needed for leptogenesis leads to an overproduction of gravitinos,
which destroys the standard predictions of Big Bang Nucleosynthesis
(BBN).
In minimal supergravity models, with $m_{3/2} \sim$ 100 GeV to 10 TeV
gravitinos are not stable, decaying during or after BBN. Their rate of
production can be so large that subsequent gravitino decays completely
change the standard BBN scenario.
To prevent such ``gravitino crisis'' one requires an upper bound on
the reheating temperature $T_R$ after inflation, since the abundance
of gravitinos is proportional to $T_R$.
A recent detailed analysis derived a stringent upper bound $T_R \lsim
10^6$ GeV when the gravitino decay has hadronic
modes~\cite{Kawasaki:2004qu}.

This upper bound is in conflict with the temperature required for
leptogenesis, $T_R > 2 \times 10^9$ GeV~\cite{Buchmuller:2004nz}.
Therefore, thermal leptogenesis seems difficult to reconcile with low
energy supersymmetry if gravitino masses lie in the range suggested by
the simplest minimal supergravity models.
Their required mass is typically too large in order for them to be
produced after inflation, implying that the minimal type I
supersymmetric seesaw schemes may be in trouble.  Two recent
suggestions have been made to cure this inconsistency. 

One proposal~\cite{Farzan:2005ez} was to add a small R-parity
violating $\lambda_i \hat{\nu^c}_i \hat{H}_u \hat{H}_d$ term in the
superpotential, where $\hat{\nu^c}_i$ are right-handed neutrino
supermultiplets. One can show that in the presence of this term, the
produced lepton-antilepton asymmetry can be enhanced.
An alternative suggestion~\cite{Hirsch:2006ft} was made in the context
of the extended supersymmetric seesaw scheme considered in
Sec.~\ref{sec:low-b-l}. It was shown in this case that leptogenesis
can occur at the TeV scale through the decay of a new singlet, thereby
avoiding the gravitino crisis. Washout of the asymmetry is effectively
suppressed by the absence of direct couplings of the singlet to
leptons.

\section{Non-standard neutrino interactions}
\label{sec:non-stand-neutr}

\begin{figure}[t] \centering
    \includegraphics[height=3.7cm,width=.5\linewidth]{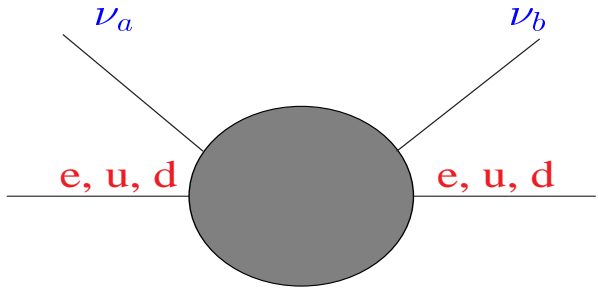}
    \caption{\label{fig:nuNSI} %
      Flavour-changing effective operator for non-standard neutrino interaction.}
\end{figure}

Most neutrino mass generation mechanisms imply the existence of
dimension-6 sub-weak strength $\varepsilon G_F$ non-standard neutrino
interaction (NSI) operators, as illustrated in Fig.~\ref{fig:nuNSI}.
These NSI can be of two types: flavour-changing (FC) and non-universal
(NU). They are conceptually interesting for neutrino propagation since
their presence leads to the possibility of resonant neutrino
conversions even in the absence of masses~\cite{valle:1987gv}.

NSI may arise from the non-trivial structure of charged and neutral
current weak interactions characterizing seesaw-type
schemes~\cite{schechter:1980gr}.
While their expected magnitude is rather model-dependent, it may well
fall within the range that will be tested in future precision
studies~\cite{Huber:2004ug}.
For example, in inverse seesaw model of Sec.~\ref{sec:inverse-seesaw}
the non-unitary piece of the lepton mixing matrix can be sizeable and
hence the induced non-standard interactions may be phenomenologically
important.  With neutrino physics entering the precision age it
becomes a challenge to scrutinize the validity of the unitary
approximation to the lepton mixing matrix, given its theoretical
fragility~\cite{schechter:1980gr}.

Relatively sizable NSI strengths may also be induced in supersymmetric
unified models~\cite{hall:1986dx} and models with radiatively induced
neutrino masses, discussed in Sec.~\ref{sec:radiative-models}.

\subsection{Atmospheric neutrinos}
\label{sec:atmosph-neutr}

The Hamiltonian describing atmospheric neutrino propagation in the
presence of NSI has, in addition to the usual oscillation part,
another term $H_\mathrm{NSI}$,
\begin{equation}
    H_\mathrm{NSI} = \pm \sqrt{2} G_F N_f
    \left( \begin{array}{cc}
        0 & \varepsilon \\ \varepsilon & \varepsilon'
    \end{array}\right) \,.
\end{equation}
Here $+(-)$ holds for neutrinos (anti-neutrinos) and $\varepsilon$ and
$\varepsilon'$ parameterize the NSI: $\sqrt{2} G_F N_f \varepsilon$ is
the forward scattering amplitude for the FC process $\nu_\mu + f \to
\nu_\tau + f$ and $\sqrt{2} G_F N_f \varepsilon'$ represents the
difference between $\nu_\mu + f$ and $\nu_\tau + f$ elastic forward
scattering (NU).  $N_f$ is the number density of the fermion $f$ along
the neutrino path.

It has been shown~\cite{fornengo:2001pm} that in such 2--neutrino
approximation, the determination of atmospheric neutrino parameters
$\Dma$ and $\sin^2\theta_\Atm$ is hardly affected by the presence of
NSI on down-type quarks ($f=d$).  Future neutrino factories will
substantially improve this bound~\cite{huber:2001zw}.

\subsection{Solar neutrinos}
\label{sec:solar-neutrinos}

Are solar oscillations robust?  Do we understand the Sun, neutrino
propagation and neutrino interactions well enough to trust current
oscillation parameter determinations?
Reactors have played a crucial role in identifying oscillations as
``the'' solution to the solar neutrino problem~\cite{pakvasa:2003zv}.

Thanks to KamLAND we have ruled out an otherwise excellent solution of
the solar neutrino problem based on spin-flavour precession due to
convective zone magnetic fields~\cite{miranda:2000bi,barranco:2002te}.
The absence of solar anti-neutrinos in KamLAND~\cite{Miranda:2004nz}
has been used to establish robustness of the LMA solution with respect
to spin-flavour precession due to convective zone magnetic fields.

Thanks to KamLAND one could also establish robustness of the LMA
solution with respect to small density fluctuations in the solar
interior~\cite{Burgess:2003su,burgess:2002we}, as could arise, say,
from solar radiative zone magnetic fields~\cite{Burgess:2003fj}.

Finally, again thanks to KamLAND we have ruled out an otherwise
excellent solution of the solar neutrino problem based on non-standard
neutrino interactions~\cite{guzzo:2001mi}.
However, in contrast to the atmospheric case, non-standard physics may
still affect neutrino propagation properties and detection cross
sections in ways that can affect current
determinations~\cite{Miranda:2004nb}.  This implies that the
oscillation interpretation of solar neutrino data is still ``fragile''
with respect to the presence of non-standard interactions in the
$e-\tau$ sector, though the required NSI strength for non-robustness
to set in is quite large.
In contrast, one can show that even a small residual non-standard
interaction of neutrinos in this channel can have dramatic
consequences for the sensitivity to $\theta_{13}$ at a neutrino
factory~\cite{huber:2001de}.  It is therefore important to improve the
sensitivities on NSI, another window of opportunity for neutrino
physics in the precision age.

\def\baselinestretch{1}
\section*{Acknowledgements}

I thank the organizers for hospitality at Corfu. This work was
supported by a Humboldt research award at the Institut f\"ur
Theoretische Physik of the Universit\"at T\"ubingen, and also by
Spanish grants FPA2005-01269/BFM2002-00345, European commission RTN
Contract MRTN-CT-2004-503369 and ILIAS/N6 Contract
RII3-CT-2004-506222. I thank T. Rashba for proof-reading.

\def\baselinestretch{1}


\begin{thebibliography}{100}

\bibitem{fukuda:2002pe}
Super-Kamiokande collaboration, S.~Fukuda {\em et~al.},
\newblock Phys. Lett. {\bf B539}, 179 (2002), [hep-ex/0205075].

\bibitem{ahmad:2002jz}
SNO collaboration, Q.~R. Ahmad {\em et~al.},
\newblock Phys. Rev. Lett. {\bf 89}, 011301 (2002), [nucl-ex/0204008].

\bibitem{araki:2004mb}
KamLAND collaboration, T.~Araki {\em et~al.},
\newblock Phys. Rev. Lett. {\bf 94}, 081801 (2004).

\bibitem{Kajita:2004ga}
T.~Kajita,
\newblock New J. Phys. {\bf 6}, 194 (2004).

\bibitem{ahn:2002up}
K2K collaboration, M.~H. Ahn {\em et~al.},
\newblock Phys. Rev. Lett. {\bf 90}, 041801 (2003), [hep-ex/0212007].

\bibitem{Minkowski:1977sc}
P.~Minkowski,
\newblock Phys. Lett. {\bf B67}, 421 (1977).

\bibitem{Orloff:2005nu} Articles by M. Gell-Mann, P. Ramond and R.
  Slansky; T.  Yanagida; R. Mohapatra and G. Senjanovic and S. Glashow
  in Proc. of Int. Conf. on the Seesaw Mechanism and the Neutrino
  Mass, Paris, France, 10-11 June 2004.  Edited by J.~Orloff,
  S.~Lavignac and M.~Cribier.

\bibitem{Weinberg:1980bf}
S.~Weinberg,
\newblock Phys. Rev. {\bf D22}, 1694 (1980).

\bibitem{schechter:1980gr}
J.~Schechter and J.~W.~F. Valle,
\newblock Phys. Rev. {\bf D22}, 2227 (1980).

\bibitem{schechter:1982cv}
J.~Schechter and J.~W.~F. Valle,
\newblock Phys. Rev. {\bf D25}, 774 (1982).

\bibitem{Lazarides:1980nt}
G.~Lazarides, Q.~Shafi and C.~Wetterich,
\newblock Nucl. Phys. {\bf B181}, 287 (1981).

\bibitem{apollonio:1999ae}
CHOOZ collaboration, M.~Apollonio {\em et~al.},
\newblock Phys. Lett. {\bf B466}, 415 (1999) 

\bibitem{boehm:2001ik}
Palo Verde collaboration, F.~Boehm {\em et~al.},
\newblock Phys. Rev. {\bf D64}, 112001 (2001) 

\bibitem{Maltoni:2004ei} M.~Maltoni, T.~Schwetz, M.~A. Tortola and
  J.~W.~F. Valle, \newblock New J. Phys. {\bf 6}, 122 (2004),
  \newblock Appendix C in hep-ph/0405172 (v5) provides updated results
  which take into account all developments as of June 2006, namely:
  new SSM, new SNO salt data, latest K2K and MINOS data; previous
  analyses e.~g. by Bahcall et al, Bandyopadhyay et al, Fogli et al,
  are referenced therein.

\bibitem{elliott:2002xe}
S.~R. Elliott and P.~Vogel,
\newblock Ann. Rev. Nucl. Part. Sci. {\bf 52}, 115 (2002) 

\bibitem{doi:1985dx}
M.~Doi, T.~Kotani and E.~Takasugi,
\newblock Prog. Theor. Phys. Suppl. {\bf 83}, 1 (1985).

\bibitem{Schechter:1982bd}
J.~Schechter and J.~W.~F. Valle,
\newblock Phys. Rev. {\bf D25}, 2951 (1982).

\bibitem{bilenky:1980cx}
S.~M. Bilenky, J.~Hosek and S.~T. Petcov,
\newblock Phys. Lett. {\bf B94} (1980) 49

\bibitem{Schechter:1981gk}
J.~Schechter and J.~W.~F. Valle,
\newblock Phys. Rev. {\bf D23}, 1666 (1981).

\bibitem{doi:1981yb}
M.~Doi, T.~Kotani, H.~Nishiura, K.~Okuda and E.~Takasugi,
\newblock Phys. Lett. {\bf B102}, 323 (1981).

\bibitem{schechter:1981hw}
J.~Schechter and J.~W.~F. Valle,
\newblock Phys. Rev. {\bf D24}, 1883 (1981),
\newblock Err. D25, 283 (1982).

\bibitem{Wolfenstein:1981rk}
L.~Wolfenstein,
\newblock Phys. Lett. {\bf B107}, 77 (1981).

\bibitem{Dienes:1998sb}
K.~R. Dienes, E.~Dudas and T.~Gherghetta,
\newblock Nucl. Phys. {\bf B557}, 25 (1999), [hep-ph/9811428].

\bibitem{Arkani-Hamed:1998vp}
N.~Arkani-Hamed, S.~Dimopoulos, G.~R. Dvali and J.~March-Russell,
\newblock Phys. Rev. {\bf D65}, 024032 (2002), [hep-ph/9811448].

\bibitem{Ioannisian:1999sw}
A.~Ioannisian and J.~W.~F. Valle,
\newblock Phys. Rev. {\bf D63}, 073002 (2001), [hep-ph/9911349].

\bibitem{deGouvea:2000jp}
A.~de~Gouvea and J.~W.~F. Valle,
\newblock Phys. Lett. {\bf B501}, 115 (2001) 

\bibitem{chikashige:1981ui}
Y.~Chikashige, R.~N. Mohapatra and R.~D. Peccei,
\newblock Phys. Lett. {\bf B98}, 265 (1981).

\bibitem{Akhmedov:2005np}
E.~K. Akhmedov and M.~Frigerio,
\newblock Phys. Rev. Lett. {\bf 96}, 061802 (2006), [hep-ph/0509299];
 P.~Hosteins, S.~Lavignac and C.~A.~Savoy,
  hep-ph/0606078.

\bibitem{mohapatra:1986bd}
R.~N. Mohapatra and J.~W.~F. Valle,
\newblock Phys. Rev. {\bf D34}, 1642 (1986).

\bibitem{Witten:1985xc}
E.~Witten,
\newblock Nucl. Phys. {\bf B258}, 75 (1985).

\bibitem{Akhmedov:1995vm}
E.~Akhmedov, M.~Lindner, E.~Schnapka and J.~W.~F. Valle,
\newblock Phys. Rev. {\bf D53}, 2752 (1996), [hep-ph/9509255];
\newblock Phys. Lett. {\bf B368}, 270 (1996), [hep-ph/9507275].

\bibitem{Barr:2005ss}
S.~M. Barr and I.~Dorsner,
\newblock Phys. Lett. {\bf B632}, 527 (2006), [hep-ph/0507067].

\bibitem{Fukuyama:2005gg}
T.~Fukuyama, A.~Ilakovac, T.~Kikuchi and K.~Matsuda,
\newblock JHEP {\bf 06}, 016 (2005), [hep-ph/0503114].

\bibitem{Malinsky:2005bi}
M.~Malinsky, J.~C. Romao and J.~W.~F. Valle,
\newblock Phys. Rev. Lett. {\bf 95}, 161801 (2005), [hep-ph/0506296].

\bibitem{'tHooft:1979bh}
G.~'t~Hooft,
\newblock Lecture given at Cargese Summer Inst., Cargese, France, Aug. 1979.

\bibitem{Joshipura:1993hp}
A.~S. Joshipura and J.~W.~F. Valle,
\newblock Nucl. Phys. {\bf B397}, 105 (1993).

\bibitem{romao:1992zx}
J.~C. Romao, F.~de~Campos and J.~W.~F. Valle,
\newblock Phys. Lett. {\bf B292}, 329 (1992), [hep-ph/9207269].

\bibitem{Hirsch:2004rw}
M.~Hirsch, {\em et~al.},
\newblock Phys. Rev. {\bf D70}, 073012 (2004), [hep-ph/0407269].

\bibitem{Hirsch:2005wd}
M.~Hirsch, {\em et~al.}, 
\newblock Phys. Rev. {\bf D73}, 055007 (2006), [hep-ph/0512257].

\bibitem{deCampos:1997bg}
F.~de~Campos, O.~J.~P. Eboli, J.~Rosiek and J.~W.~F. Valle,
\newblock Phys. Rev. {\bf D55}, 1316 (1997), [hep-ph/9601269].

\bibitem{Abdallah:2003ry}
DELPHI collaboration, J.~Abdallah {\em et~al.},
\newblock Eur. Phys. J. {\bf C32}, 475 (2004), [hep-ex/0401022].

\bibitem{zee:1980ai}
A.~Zee,
\newblock Phys. Lett. {\bf B93}, 389 (1980).

\bibitem{babu:1988ki}
K.~S. Babu,
\newblock Phys. Lett. {\bf B203}, 132 (1988).

\bibitem{Peltoniemi:1993pd}
J.~T. Peltoniemi and J.~W.~F. Valle,
\newblock Phys. Lett. {\bf B304}, 147 (1993), [hep-ph/9301231].

\bibitem{Hirsch:2004he}
M.~Hirsch and J.~W.~F. Valle,
\newblock New J. Phys. {\bf 6}, 76 (2004), [hep-ph/0405015].

\bibitem{Masiero:1990uj}
A.~Masiero and J.~W.~F. Valle,
\newblock Phys. Lett. {\bf B251}, 273 (1990).

\bibitem{romao:1992vu}
J.~C. Romao, C.~A. Santos and J.~W.~F. Valle,
\newblock Phys. Lett. {\bf B288}, 311 (1992).

\bibitem{romao:1997xf}
J.~C. Romao, A.~Ioannisian and J.~W.~F. Valle,
\newblock Phys. Rev. {\bf D55}, 427 (1997), [hep-ph/9607401].

\bibitem{Diaz:1998xc}
M.~A. Diaz, J.~C. Romao and J.~W.~F. Valle,
\newblock Nucl. Phys. {\bf B524}, 23 (1998), [hep-ph/9706315].

\bibitem{Hirsch:2000ef}
M.~Hirsch {\em et~al.},
\newblock Phys. Rev. {\bf D62}, 113008 (2000), [hep-ph/0004115],
\newblock Err-ibid. {\bf D65}:119901,2002;
M.~A.~Diaz {\em et~al.},
  Phys.\ Rev.\ D {\bf 68} (2003) 013009
  [Erratum-ibid.\ D {\bf 71} (2005) 059904]
  [hep-ph/0302021].

\bibitem{gonzalez-garcia:1989rw}
M.~C. Gonzalez-Garcia and J.~W.~F. Valle,
\newblock Phys. Lett. {\bf B216}, 360 (1989).

\bibitem{branco:1989bn}
G.~C. Branco, M.~N. Rebelo and J.~W.~F. Valle,
\newblock Phys. Lett. {\bf B225}, 385 (1989).

\bibitem{rius:1990gk}
N.~Rius and J.~W.~F. Valle,
\newblock Phys. Lett. {\bf B246}, 249 (1990).

\bibitem{Bahcall:2004fg}
J.~N. Bahcall and M.~H. Pinsonneault,
\newblock Phys. Rev. Lett. {\bf 93}, 121301 (2004), [astro-ph/0402114].

\bibitem{Honda:2004yz}
M.~Honda, T.~Kajita, K.~Kasahara and S.~Midorikawa,
\newblock Phys.\ Rev.\ D {\bf 70} (2004) 043008
  [astro-ph/0404457].

\bibitem{mikheev:1985gs}
S.~P. Mikheev and A.~Y. Smirnov,
\newblock Sov. J. Nucl. Phys. {\bf 42}, 913 (1985).

\bibitem{wolfenstein:1978ue}
L.~Wolfenstein,
\newblock Phys. Rev. {\bf D17}, 2369 (1978).

\bibitem{Bahcall:2005va}
J.~N. Bahcall, A.~M. Serenelli and S.~Basu,
\newblock astro-ph/0511337.

\bibitem{Aharmim:2005gt}
SNO collaboration, B.~Aharmim {\em et~al.},
\newblock Phys. Rev. {\bf C72}, 055502 (2005), [nucl-ex/0502021].

\bibitem{Ahn:2006zz}
K2K collaboration, M.~H. Ahn,
\newblock hep-ex/0606032.

\bibitem{Tagg:2006sx} MINOS colaboration, N.~Tagg, \newblock
  hep-ex/0605058\\ see also
  http://www-numi.fnal.gov/talks/results06.html

\bibitem{pakvasa:2003zv}
S.~Pakvasa and J.~W.~F. Valle,
\newblock hep-ph/0301061,
\newblock Proc. of the Indian National Academy of Sciences on Neutrinos, Vol.
  70A, No.1, p.189 - 222 (2004), Eds. D. Indumathi, M.V.N. Murthy and G.
  Rajasekaran.

\bibitem{gonzalez-garcia:2000sq}
M.~C. Gonzalez-Garcia {\em et~al.},
\newblock Phys. Rev. {\bf D63}, 033005 (2001), [hep-ph/0009350].

\bibitem{schechter:1980bn}
J.~Schechter and J.~W.~F. Valle,
\newblock Phys. Rev. {\bf D21}, 309 (1980).

\bibitem{Alsharoa:2002wu}
Muon Collider/Neutrino Factory, M.~M. Alsharoa {\em et~al.},
\newblock Phys. Rev. ST Accel. Beams {\bf 6}, 081001 (2003), [hep-ex/0207031].

\bibitem{apollonio:2002en}
M.~Apollonio {\em et~al.},
\newblock CERN Yellow Report on the Neutrino Factory,
\newblock hep-ph/0210192.

\bibitem{albright:2000xi}
C.~Albright {\em et~al.},
\newblock Report to the Fermilab Directorate,
\newblock hep-ex/0008064.

\bibitem{Huber:2004ug}
P.~Huber, M.~Lindner, M.~Rolinec, T.~Schwetz and W.~Winter,
\newblock Phys. Rev. {\bf D70}, 073014 (2004), [hep-ph/0403068].

\bibitem{SKatm04}
C.~Yanagisawa,
\newblock Proc. of International Workshop on Astroparticle and High Energy
  Physics, October 14 - 18, 2003, Valencia, Spain, published at JHEP,
  PRHEP-AHEP2003/062, accessible from http://ahep.uv.es/.

\bibitem{Akhmedov:2004rq}
E.~K. Akhmedov, M.~A. Tortola and J.~W.~F. Valle,
\newblock JHEP {\bf 05}, 057 (2004), [hep-ph/0404083].

\bibitem{Raidal:2004iw}
M.~Raidal,
\newblock Phys. Rev. Lett. {\bf 93}, 161801 (2004), [hep-ph/0404046].

\bibitem{minakata-2004-70}
H.~Minakata and A.~Y. Smirnov,
\newblock Physical Review {\bf D70}, 073009 (2004).

\bibitem{Ferrandis:2004vp}
J.~Ferrandis and S.~Pakvasa,
\newblock Phys. Rev. {\bf D71}, 033004 (2005), [hep-ph/0412038].

\bibitem{Dighe:2006zk}
A.~Dighe, S.~Goswami and P.~Roy,
\newblock Phys. Rev. {\bf D73}, 071301 (2006) 

\bibitem{Harrison:2002kp}
P.~F. Harrison and W.~G. Scott,
\newblock Phys. Lett. {\bf B535}, 163 (2002), [hep-ph/0203209].

\bibitem{Harrison:2002er}
P.~F. Harrison, D.~H. Perkins and W.~G. Scott,
\newblock Phys. Lett. {\bf B530}, 167 (2002), [hep-ph/0202074].

\bibitem{chankowski:2000fp}
P.~Chankowski, A.~Ioannisian, S.~Pokorski and J.~W.~F. Valle,
\newblock Phys. Rev. Lett. {\bf 86}, 3488 (2001), [hep-ph/0011150].

\bibitem{amaldi:1991cn}
U.~Amaldi, W.~de~Boer and H.~Furstenau,
\newblock Phys. Lett. {\bf B260}, 447 (1991).

\bibitem{babu:2002dz}
K.~S. Babu, E.~Ma and J.~W.~F. Valle,
\newblock Phys. Lett. {\bf B552}, 207 (2003), [hep-ph/0206292].

\bibitem{Hirsch:2005mc}
M.~Hirsch, {\em et~al.},
\newblock Phys. Rev. {\bf D72}, 091301 (2005), [hep-ph/0507148].

\bibitem{Grimus:2003yn}
W.~Grimus and L.~Lavoura,
\newblock Phys. Lett. {\bf B579}, 113 (2004), [hep-ph/0305309].

\bibitem{Altarelli:2005yp}
G.~Altarelli and F.~Feruglio,
\newblock Nucl.\ Phys.\ B {\bf 720} (2005) 64 [hep-ph/0504165].

\bibitem{Hirsch:2006je}
M.~Hirsch, {\em et~al.},
\newblock hep-ph/0606082.

\bibitem{Altarelli:2004za}
G.~Altarelli and F.~Feruglio,
\newblock New J. Phys. {\bf 6}, 106 (2004), [hep-ph/0405048].

\bibitem{pal:1982rm}
P.~B. Pal and L.~Wolfenstein,
\newblock Phys. Rev. {\bf D25}, 766 (1982).

\bibitem{kayser:1982br}
B.~Kayser,
\newblock Phys. Rev. {\bf D26}, 1662 (1982).

\bibitem{Drexlin:2005zt}
KATRIN collaboration, G.~Drexlin,
\newblock Nucl. Phys. Proc. Suppl. {\bf 145}, 263 (2005).

\bibitem{Lesgourgues:2006nd}
J.~Lesgourgues and S.~Pastor,
\newblock Phys. Rep. {\bf 429}, 307 (2006), [astro-ph/0603494].

\bibitem{Hannestad:2006zg}
S.~Hannestad,
\newblock hep-ph/0602058.

\bibitem{Fogli:2004as} G.~L.~Fogli, E.~Lisi, A.~Marrone,
  A.~Melchiorri, A.~Palazzo, P.~Serra and J.~Silk, \newblock
  Phys.\ Rev.\ D {\bf 70} (2004) 113003 

\bibitem{Bilenky:2004wn}
S.~M. Bilenky, A.~Faessler and F.~Simkovic,
\newblock Phys. Rev. {\bf D70}, 033003 (2004), [hep-ph/0402250].

\bibitem{caldwell:1993kn}
D.~O. Caldwell and R.~N. Mohapatra,
\newblock Phys. Rev. {\bf D48}, 3259 (1993).

\bibitem{ioannisian:1994nx}
A.~Ioannisian and J.~W.~F. Valle,
\newblock Phys. Lett. {\bf B332}, 93 (1994), [hep-ph/9402333].

\bibitem{Klapdor-Kleingrothaus:2004wj}
H.~V. Klapdor-Kleingrothaus, I.~V. Krivosheina, A.~Dietz and O.~Chkvorets,
\newblock Phys. Lett. {\bf B586}, 198 (2004), [hep-ph/0404088].

\bibitem{Aalseth:2002dt}
C.~E. Aalseth {\em et~al.},
\newblock Mod. Phys. Lett. {\bf A17}, 1475 (2002), [hep-ex/0202018].

\bibitem{Aalseth:2002rf}
IGEX collaboration, C.~E. Aalseth {\em et~al.},
\newblock Phys. Rev. {\bf D65}, 092007 (2002), [hep-ex/0202026].

\bibitem{dbd06}
R.~Saakyan, C.~Nones, C.~Tomei and K.~Zuber,
\newblock DBD06 - ILIAS/N6 WG1 Collaboration meeting, April, 2006, accessible
  from http://ahep.uv.es/dbd06/index.php.

\bibitem{kachelriess:2000qc}
M.~Kachelriess, R.~Tomas and J.~W.~F. Valle,
\newblock Phys. Rev. {\bf D62}, 023004 (2000), [hep-ph/0001039].

\bibitem{valle:1987sq}
J.~W.~F. Valle,
\newblock Phys. Lett. {\bf B196}, 157 (1987).

\bibitem{hall:1986dx}
L.~J. Hall, V.~A. Kostelecky and S.~Raby,
\newblock Nucl. Phys. {\bf B267}, 415 (1986).

\bibitem{borzumati:1986qx}
F.~Borzumati and A.~Masiero,
\newblock Phys. Rev. Lett. {\bf 57}, 961 (1986).

\bibitem{casas:2001sr}
J.~A. Casas and A.~Ibarra,
\newblock Nucl. Phys. {\bf B618}, 171 (2001), [hep-ph/0103065].
 S.~Antusch, E.~Arganda, M.~J.~Herrero and A.~Teixeira,
 hep-ph/0607263.

\bibitem{Deppisch:2004fa}
F.~Deppisch and J.~W.~F. Valle,
\newblock Phys. Rev. {\bf D72}, 036001 (2005), [hep-ph/0406040].

\bibitem{bernabeu:1987gr}
J.~Bernabeu {\em et~al.},
\newblock Phys. Lett. {\bf B187}, 303 (1987).

\bibitem{gonzalez-garcia:1992be}
M.~C. Gonzalez-Garcia and J.~W.~F. Valle,
\newblock Mod. Phys. Lett. {\bf A7}, 477 (1992).

\bibitem{Ilakovac:1994kj}
A.~Ilakovac and A.~Pilaftsis,
\newblock Nucl. Phys. {\bf B437}, 491 (1995), [hep-ph/9403398].

\bibitem{Dittmar:1990yg}
M.~Dittmar {\em et~al.},
\newblock Nucl. Phys. {\bf B332}, 1 (1990);
DELPHI collaboration, P.~Abreu {\em et~al.},
\newblock Z. Phys. {\bf C74}, 57 (1997).

\bibitem{Deppisch:2005zm}
F.~Deppisch, T.~S. Kosmas and J.~W.~F. Valle,
\newblock hep-ph/0512360.

\bibitem{Kuno:2000kd}
Y.~Kuno,
\newblock AIP Conf. Proc. {\bf 542}, 220 (2000).

\bibitem{deCampos:2005ri}
F.~de~Campos {\em et~al.},
\newblock Phys. Rev. {\bf D71}, 075001 (2005), [hep-ph/0501153].

\bibitem{Porod:2000hv}
W.~Porod, {\em et~al.},
\newblock Phys. Rev. {\bf D63}, 115004 (2001), [hep-ph/0011248].

\bibitem{romao:1999up}
J.~C. Romao {\em et~al.},
\newblock Phys. Rev. {\bf D61}, 071703 (2000), [hep-ph/9907499].

\bibitem{mukhopadhyaya:1998xj}
B.~Mukhopadhyaya, S.~Roy and F.~Vissani,
\newblock Phys. Lett. {\bf B443}, 191 (1998), [hep-ph/9808265].

\bibitem{Hirsch:2003fe}
M.~Hirsch and W.~Porod,
\newblock Phys. Rev. {\bf D68}, 115007 (2003), [hep-ph/0307364].

\bibitem{Ma:2000xh}
  E.~Ma, M.~Raidal and U.~Sarkar,
  Nucl.\ Phys.\ B {\bf 615} (2001) 313
  [hep-ph/0012101].

\bibitem{Fukugita:1986hr}
M.~Fukugita and T.~Yanagida,
\newblock Phys. Lett. {\bf B174}, 45 (1986).

\bibitem{Buchmuller:2005eh}
W.~Buchmuller, R.~D. Peccei and T.~Yanagida,
\newblock Ann. Rev. Nucl. Part. Sci. {\bf 55}, 311 (2005), [hep-ph/0502169].

\bibitem{kuzmin:1985mm}
V.~A. Kuzmin, V.~A. Rubakov and M.~E. Shaposhnikov,
\newblock Phys. Lett. {\bf B155}, 36 (1985).

\bibitem{Kawasaki:2004qu}
M.~Kawasaki, K.~Kohri and T.~Moroi,
\newblock Phys. Rev. {\bf D71}, 083502 (2005) 

\bibitem{Buchmuller:2004nz}
W.~Buchmuller, P.~Di~Bari and M.~Plumacher,
\newblock Annals Phys.\  {\bf 315} (2005) 305

\bibitem{Farzan:2005ez}
Y.~Farzan and J.~W.~F. Valle,
\newblock Phys. Rev. Lett. {\bf 96}, 011601 (2006), [hep-ph/0509280].

\bibitem{Hirsch:2006ft}
M.~Hirsch, M.~Malinsky, J.~C. Romao, U.~Sarkar and J.~W.~F. Valle,
\newblock hep-ph/0608006.

\bibitem{valle:1987gv}
J.~W.~F. Valle,
\newblock Phys. Lett. {\bf B199}, 432 (1987).

\bibitem{fornengo:2001pm}
N.~Fornengo {\em et~al.},
\newblock Phys. Rev. {\bf D65}, 013010 (2002), [hep-ph/0108043].

\bibitem{huber:2001zw}
P.~Huber and J.~W.~F. Valle,
\newblock Phys. Lett. {\bf B523}, 151 (2001), [hep-ph/0108193].

\bibitem{miranda:2000bi}
O.~G. Miranda {\em et~al.},
\newblock Nucl. Phys. {\bf B595}, 360 (2001), [hep-ph/0005259];
\newblock Phys. Lett. {\bf B521}, 299 (2001), [hep-ph/0108145].

\bibitem{barranco:2002te}
J.~Barranco, {\em et~al.},
\newblock Phys. Rev. {\bf D66}, 093009 (2002), [hep-ph/0207326]

\bibitem{Miranda:2004nz}
O.~G. Miranda, T.~I. Rashba, A.~I. Rez and J.~W.~F. Valle,
\newblock Phys. Rev. {\bf D70}, 113002 (2004), [hep-ph/0406066];
\newblock Phys. Rev. Lett. {\bf 93}, 051304 (2004), [hep-ph/0311014].

\bibitem{Burgess:2003su}
C.~P. Burgess {\em et~al.},
\newblock JCAP {\bf 0401}, 007 (2004), [hep-ph/0310366].

\bibitem{burgess:2002we}
C.~Burgess {\em et~al.},
\newblock Astrophys. J. {\bf 588}, L65 (2003), [hep-ph/0209094].

\bibitem{Burgess:2003fj}
C.~P. Burgess, {\em et~al.},
\newblock Mon. Not. Roy. Astron. Soc. {\bf 348}, 609 (2004),
  [astro-ph/0304462].

\bibitem{guzzo:2001mi}
M.~Guzzo {\em et~al.},
\newblock Nucl. Phys. {\bf B629}, 479 (2002), [hep-ph/0112310 v3
  KamLAND-updated version].

\bibitem{Miranda:2004nb}
O.~G. Miranda, M.~A. Tortola and J.~W.~F. Valle,
\newblock hep-ph/0406280.

\bibitem{huber:2001de}
P.~Huber, T.~Schwetz and J.~W.~F. Valle,
\newblock Phys. Rev. Lett. {\bf 88}, 101804 (2002), [hep-ph/0111224];
\newblock Phys. Rev. {\bf D66}, 013006 (2002), [hep-ph/0202048].

\end{thebibliography}
\end{document}